\documentclass[
  preprint,
  nofootinbib,
  amsmath,amssymb,
  aps
]{revtex4-2}

\usepackage{graphicx}
\usepackage{dcolumn}
\usepackage{bm}
\usepackage{rotating}
\usepackage{newtxtext,newtxmath,tablefootnote,multirow}
\usepackage[T1]{fontenc}
\usepackage{tabularx}
\usepackage[normalem]{ulem}
\usepackage{amsmath}
\usepackage{diagbox}
\usepackage{xcolor}

\useunder{\uline}{\ul}{}

\begin{document}


\title{\textbf{Probing Gravitational Wave Speed and Dispersion with LISA Observations of Supermassive Black Hole Binary Populations}
}%

\author{Tian-Yong Cao$^{1,2}$}
\author{Shu-Xu Yi$^{1}$}%
 \email{Contact author: sxyi@ihep.ac.cn}
\address{$^{1}$Key Laboratory of Particle Astrophysics, Institute of High Energy Physics, Chinese Academy of Sciences, 19B Yuquan Road, Beijing 100049, People’s Republic of China}
\address{$^{2}$University of Chinese Academy of Sciences, Chinese Academy of Sciences, Beijing 100049, People’s Republic of China}%



\date{October 30, 2025}

\begin{abstract}
\noindent According to General Relativity (GR), gravitational waves (GWs) should travel at the speed of light $c$. However, some theories beyond GR predict deviations of the velocity of GWs $c_{\rm gw}$ from $c$, and some of those expect vacuum dispersion. Therefore, probing the propagation effects of GWs by comparing the wave format
detectors against the one at emission excepted from GR. Since such propagation effects accumulate through larger distance, it is expected that super-massive black holes binary (SMBHB) mergers serve as better targets than their stellarmass equivalent. In this paper, we study with simulations on how observations on a population of SMBHs can help to study this topic. We simulate LISA observations on three possible SMBHB merger populations, namely Pop\MakeUppercase{\romannumeral 3}, Q3-nod and Q3-d over a 5-year mission. The resulting constraints on the graviton mass are \(9.50\), \(9.33\), and \(9.05 \times 10^{-27} \, \mathrm{eV}/c^2\), respectively. We also obtain the corresponding constraints on the dispersion coefficients assuming different dispersion scenarios. If the electromagnetic wave counterparts of SMBHB merger can be detected simultaneously, the $c_{\rm gw}$ can be constrained waveform-independently to \(\Delta c/c\) to \(10^{-13}-10^{-12}\), corresponding to graviton mass constraints of \(10^{-26}-10^{-24} \mathrm{eV}/c^2\).
\end{abstract}

\maketitle


\section{Introduction}

\label{introduction}

General Relativity (GR) is the most successful theory of gravity to date and has been subjected to various tests over the past century. With the advent of the gravitational wave (GW) and multimessenger era, continuous observations of GWs have been employed to test the GR \citep{abbott2016observation, abbott2016tests}. These results consistently show that, with increasing precision, GR remains the best theory for describing gravitation.

Motivated by various theoretical considerations, many theories of gravity beyond GR have been proposed to address issues that arise in the quest for a more complete understanding of the fundamental forces. Among these challenges are potential deviations from Lorentz invariance, which could emerge in frameworks attempting to unify quantum mechanics and gravity. For instance \citep{will2014confrontation}, quantum gravity models introducing a minimal length scale, such as the Planck length, suggest spacetime may exhibit granular properties, potentially leading to Lorentz violations at high energies. In brane-world scenarios, where our universe is a four-dimensional "brane" in a higher-dimensional bulk, the projection of physics onto the brane might induce apparent Lorentz-violating effects. Similarly, string theory and other unified frameworks often include additional fields that couple to Standard Model particles, potentially leading to effective Lorentz violations. Alternative theories of gravity, such as scalar-tensor or massive gravity, also introduce modifications to spacetime that could challenge Lorentz invariance.

Experimentally, the difference between the measured speed of light \(c\) and the theoretical speed limit \(c_0\) can be used to constrain these theories. Colladay \& Kosteleck{\`y}; Kosteleck{\`y} \& Mewes \cite{colladay1997cpt, colladay1998lorentz, kostelecky2002signals} extended these considerations to the entire Standard Model of particle physics, forming the framework known as the Standard Model Extension (SME), which systematically characterizes deviations between \(c\) and \(c_0\) \citep{Kostelecky2016Testing,Mewes2019Signals}:
\begin{equation}
    c = 1-\varsigma^{0(d)}(f)\pm|\vec{\varsigma}^{(d)}|(f)\,,
\end{equation}
Here we let \(h = c_0 = G = 1\), the same as below. When the spacetime dimension is \( d = 4 \), \(\varsigma^{0(4)}\) and \(|\vec{\varsigma}^{(4)}|\) are independent of the frequency of light. However, when the dimension is extended to \( d = 5,6 \), dispersion effects begin to emerge. 

The results of Lorentz invariance are not limited to light but also apply to GWs, meaning that the speed of GWs, \( c_{\rm GW} \), follows the same relationship with \( c_0 \) as described in the equation above. Moreover, more specific theories predicting GW dispersion have also been proposed. For instance, Double Special Relativity (DSR) introduces modifications to traditional relativity, particularly in the high-energy regime near the Planck scale, where both quantum mechanics and GR are expected to have significant effects \citep{amelino2001testable,magueijo2002lorentz,amelino2002special,amelino2010doubly}. Moreover, the existence of a massive graviton is a key premise in many approaches that aim to reconcile the discrepancies between high-energy GR and quantum mechanics. A direct consequence of such theories is the modification of the thermodynamics of a FRW universe, which can be fully described using the generalized uncertainty principle \citep{sefiedgar2011modified}, and modifications in dispersion relations, as explored in Extra-Dimensional Theories (ED) \citep{sefiedgar2011modified}, which include extra spatial dimensions. The Ho{\v{r}}ava-Lifshitz theory (HL) \citep{hovrava2009membranes,hovrava2009quantum,vacaru2012modified}, on the other hand, tackles quantum gravity challenges by introducing anisotropic scaling in spacetime, aiming to address the issues of quantum gravity without requiring explicit quantum gravity effects at low energies. Lastly, theories based on Non-Commutative Geometries (NCG) \citep{garattini2011modified,garattini2012modified,garattini2012particle} provide an intriguing approach to unifying GR and quantum mechanics by allowing spacetime to exhibit non-commutative properties at the Planck scale.

Generally, the dispersion relation for GWs can be written as:
\begin{equation}  
    E^2 = p^2 + \mathbb{A}_\alpha p^\alpha\,. 
\end{equation}  
In the case when \(\alpha = 0\) and $\mathbb{A}_0$>0, it is equivalent to a massive graviton scenario \(\mathbb{A}_0 = m_g^2\). Different theories predict different values for \(\alpha\) and \(\mathbb{A}_\alpha\), which are shown in Table \ref{theory}.   

\begin{table}
\centering
\caption{\textbf{The Table of \(\alpha\) and \(\mathbb{A}_\alpha\) in the dispersion relation under different theories, along with their physical significance.}}
\setlength{\tabcolsep}{40pt}
\begin{tabular}{ccc}
\hline
Theory                      & \(\alpha\)   & \(\mathbb{A}_\alpha\) \\ \hline
Standard Model Extension    & 3,4 &  \(-2\varsigma^{(5)0},-2\varsigma^{(6)0}\)  \\ 
Double Special Relativity   & 0,3 &  \(m_g,\eta_{\rm dsr}\)  \\ 
Extra-Dimensional Theories  & 0,4 &  \(m_g,-\alpha_{\rm ed}\)  \\ 
Ho{\v{r}}ava-Lifshitz theory & 4   &  \(k_{\rm hl}^4\mu_{\rm hl}^2/16\)  \\ 
Non-Commutative Geometries  & 0,4 & \(m_g,2\alpha_{\rm ncg}^2/E_p^2\)  \\ \hline 
\end{tabular}
\label{theory}
\end{table}

\begin{figure}
    \centering
    \includegraphics[width=.8\linewidth]{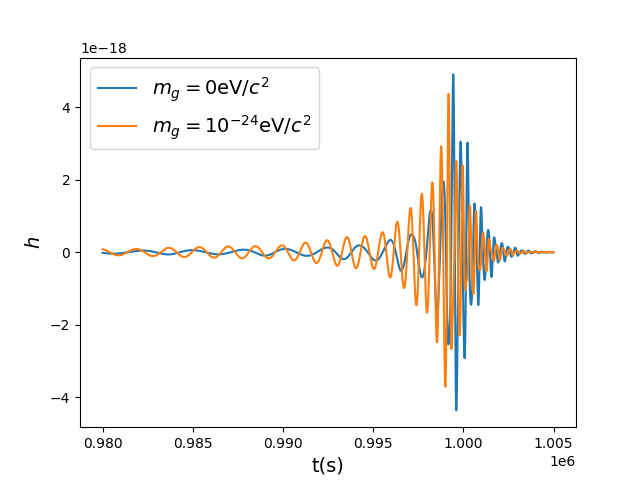}
    \caption{\textbf{A comparison of the waveforms received by LISA with and without the graviton mass}: The horizontal axis represents time, assuming the burst occurs at \(10^6\) s. The vertical axis represents the response detected by one of LISA's interferometers. The graviton mass is assumed to be \(m_g = 10^{-24} \mathrm{eV}/c^2\), with other source parameters given in Table \ref{value1570}.} 
    \label{h_time}
\end{figure}

In the case of binary black hole mergers, the GW amplitude and frequency typically rise sharply over a short duration in the inspiral and merger phases, and decay exponentially during the ring down phase. If gravitons possess a nonzero mass, dispersion causes the high-frequency GW components in the waveform arrive earlier than the low-frequency components, leading to a compression of the early waveform and a stretching of the later part (Figure \ref{h_time}). Such distortion of the GW waveform due to dispersion can be quantitatively described as follows.

Denoting the GW waveform as:
\begin{equation}
    h(t) = A(t)e^{-i\Phi(t)}\,,
\end{equation}
and applying the stationary phase approximation (SPA), its Fourier transform can be approximated as \cite{cutler1994gravitational}:
\begin{equation}
    \tilde{h}(\tilde{f}) = \frac{\tilde{A}(\tilde{t})}{\sqrt{\dot{f}(\tilde{t})}}e^{i\Psi(\tilde{f})}\,,
\end{equation}
where \(\tilde{f} = f(\tilde{t})\) is the GW frequency at the detector, and \(\tilde{h}\), \(\tilde{A}\) and \(\tilde{t}\) represent the Fourier-transformed GW strain, amplitude and time, respectively, with \(\Psi\) being the Fourier-transformed phase. The SPA assumes that the amplitude \(A(t)\) varies slowly compared to the phase \(\Phi(t)\), i.e., \(|\dot{A}/A| \ll |\dot{\Phi}|\), and that \(\Phi(t)\) is a smooth and rapidly varying function of time, so that the Fourier integral is dominated by contributions near stationary points satisfying \(\dot{\Phi}(t) = 2\pi f\).

Mirshekari et al. 
 \cite{mirshekari2012constraining} calculated that when accounting for the modified dispersion relation, we have:
\begin{equation}
    \Psi(\tilde{f}) = \Psi_{\rm GR} + \delta\Psi(\tilde{f})\,,
\end{equation}
where \(\Psi_{\rm GR}\) is the phase calculated according to GR, while \(\delta\Psi(\tilde{f})\) is given by:
\begin{equation}
    \delta\Psi(f)=
    \begin{cases} 
    -\zeta u^{\alpha-1}, &\alpha \neq 1; \\
    \zeta \ln{u}, &\alpha = 1. 
    \end{cases}
    \label{phi_zeta}
\end{equation}
with \(u = \pi \mathcal{M} f\), where \(\mathcal{M}\) denotes the chirp mass. Furthermore, we have:
\begin{equation}
    \zeta=\begin{cases}
        \frac{\pi^{2-\alpha}}{(1-\alpha)}\frac{D_\alpha}{\lambda_\mathbb{A}^{2-\alpha}}\frac{\mathcal{M}^{1-\alpha}}{(1+Z)^{1-\alpha}}, &\alpha\neq1\,;\\
        \frac{\pi D_1}{\lambda_\mathbb{A}}, &\alpha=1\,,
        \end{cases}
    \label{zeta}
\end{equation}
where \(\lambda_\mathbb{A} = |\mathbb{A}_\alpha|^{1/(\alpha-2)}\), and \(D_\alpha\) is defined by Mirshekari et al. \cite{mirshekari2012constraining}:
\begin{equation}
    D_\alpha \equiv \big({1+z\over a_0}\big)^{1-\alpha}\int_{t_e}^{t_a}a(t)^{1-\alpha}\mathrm{d}t\,,
    \label{D_alpha}
\end{equation}
where \(a_0=a(t_a)\) is the present value of the scale factor.

Specifically, when \(\mathbb{A}_\alpha\)is associated with the graviton mass, {\it i.e.,} for \(\alpha=0\) while \(\mathbb{A}_0=m_g^2\), the phase shift is given by:
\begin{equation}
    \delta\Psi(f) = -\beta_0 u^{-1}\,,
    \label{phi_mg}
\end{equation}
where \(\beta_0=\frac{\pi^2D_0\mathcal{M}}{\lambda_g^2(1+z)}\), \(D_0\) is \(D_\alpha\) when \(\alpha=0\). By fitting the observed signal phase to the phase predicted by the above equation, one can constrain the graviton mass. For instance, using 43 events from GWTC-3 \citep{abbott2021tests}, the LIGO team constrained the graviton mass to \(m_g\le1.27\times10^{-23}\,\mathrm{eV}/c^2\). Similarly, other \(\mathbb{A}_\alpha\) parameters can be constrained. The constraints for \(\alpha=0,0.5,1,1.5,2.5,3,3.5,4\) are also presented in Table \ref{result_0} of Abbott et al. \cite{abbott2021tests}. 

\begin{table*}
\centering
\caption{\textbf{The dispersion parameter constraints obtained using the sources listed in Table \ref{value1570}.}}
\setlength{\tabcolsep}{3pt}
\label{result1570}
\begin{tabular}{ccccccccccccccccc}
\hline
\multirow{2}{*}{\begin{tabular}[c]{@{}c@{}}\(m_g(\mathrm{eV}/c^2)\)\\ (\(10^{-27}\)) \end{tabular}} & \multicolumn{2}{c}{\begin{tabular}[c]{@{}c@{}}\(\mathbb{A}_{0}\)\\ (\(10^{-53}\)) \end{tabular}} & \multicolumn{2}{c}{\begin{tabular}[c]{@{}c@{}}\(\mathbb{A}_{0.5}\)\\ (\(10^{-44}\))\end{tabular}} & \multicolumn{2}{c}{\begin{tabular}[c]{@{}c@{}}\(\mathbb{A}_{1}\)\\ (\(10^{-35}\))\end{tabular}} & \multicolumn{2}{c}{\begin{tabular}[c]{@{}c@{}}\(\mathbb{A}_{1.5}\)\\ (\(10^{-27}\))\end{tabular}} & \multicolumn{2}{c}{\begin{tabular}[c]{@{}c@{}}\(\mathbb{A}_{2.5}\)\\ (\(10^{-10}\))\end{tabular}} & \multicolumn{2}{c}{\begin{tabular}[c]{@{}c@{}}\(\mathbb{A}_{3}\)\\ (\(10^{-2}\))\end{tabular}} & \multicolumn{2}{c}{\begin{tabular}[c]{@{}c@{}}\(\mathbb{A}_{3.5}\)\\ (\(10^{7}\))\end{tabular}} & \multicolumn{2}{c}{\begin{tabular}[c]{@{}c@{}}\(\mathbb{A}_{4}\)\\ (\(10^{15}\))\end{tabular}} \\ \cline{2-17}
& + & - & + & - & + & - & + & - & + & - & + & - & + & - & + & - \\ \hline
 6.79 & 4.62 & 4.76 & 2.44 & 2.46 & 1.30 & 1.32 & 8.05 & 8.27 & 6.99 & 7.05 & 9.94 & 9.67 & 9.01 & 1.76 & 3.43 & 3.39 \\ \hline
\end{tabular}
\label{result_0}
\end{table*}

One disadvantage of the above mentioned waveform based method is that their validity rely on the accuracy of the theoretical waveform at the source. Once the waveform at the source is altered from the templates we are using, due to post Newtonian corrections or environment effects, the constraints are no longer reliable. Another issue is that the above mentioned method can only be used to place constraints on theories which expect dispersion of GWs. While there are still theories predict $c_{\rm gw}\neq c$ but without GW dispersion, {\it e.g., } in the SME framework where the mass dimension equals to 4. If the electromagnetic waves (EMW) simultaneously emitted can be detected along with the GWs, their time delay can be used to place constraints on the theories, which is independent of the waveform templates and does not rely on GW dispersion.

On such observational constraint comes from GW170817 and its electromagnetic counterpart, GRB170817A. According to joint observations by LIGO and the Fermi Laboratory, GRB170817A lagged behind GW170817 by \(1.74\pm0.05\,\mathrm{s}\) \citep{abbott2017gw170817}. Assuming that the time difference \(\Delta t\) between the two signals consists of the difference in their emission times \(\Delta t_e\) and the propagation-induced time difference \(\Delta t_t\), where the latter arises due to differences in their propagation speeds:
\begin{equation}
    \begin{aligned}
            \Delta t & = \Delta t_e + \Delta t_t \\
            & = \Delta t_{e} + \frac{d_L}{c} - \frac{d_L}{c_{\mathrm{GW}}} \\
            & = \Delta t_{e} - d_L\frac{c - c_{\mathrm{GW}}}{c^2} \\
            & = \Delta t_{e} - d_L\frac{\Delta c}{c^2}\,,
    \end{aligned}
    \label{time_diff}
\end{equation}
where \(d_L\) is the luminosity distance of the source and \(\Delta c=c - c_{\mathrm{GW}}\) is the diffience of speed. Assuming that the gamma-ray burst occurred within a 0–10 s window after the binary merger, {\it i.e.}, the GW emission time, the speed of GWs can be constrained to \(-3\times10^{-15}\le\Delta c/c\le+7\times10^{-16}\) \citep{abbott2017gravitational}. Rao et al. \cite{rao2024simulation} simulated the constraints from observations on a population of Binary Neutron Star (BNS) mergers with different future GW detectors, pushing the limits on \(\Delta c/c\) to $10^{-17}$ with aLIGO and $10^{-18}$ with Einstein telescope.  

Besides of the BNS mergers, the multi-messenger observation on galactic white dwarf binaries can also be employed to place such limits \citep{cao2024constraining}.

In both the GW waveform comparison and the time difference of multi-messengers ways, one will expect more prominent propagation effects come from farther sources. In turn, the observations on sources with larger distance can result in more stringent constraints.

The hierarchical structure formation theory predicts that galaxies have undergone numerous mergers \citep{volonteri2003assembly,shen2009supermassive}, suggesting the widespread existence of SMBHBs in the universe. Recently, International Pulsar Timing Arrays (PTA) published a series of papers \citep{agazie2023nanograv,antoniadis2023second,reardon2023search,xu2023searching} providing evidence for the existence of a GW background (GWB) \footnote{CPTA reports 4.6\(\sigma\) evidence for Hellings-Downs (HD) spatial correlations indicative of a nanohertz GWB at 14 nHz. EPTA finds marginal GWB evidence in its full dataset (Bayes factor 4, 4\% false-alarm probability) and stronger evidence in a subset (Bayes factor 60, \(\sim0.1\%\) false-alarm or \(\sim3\sigma\)), characterized by HD angular patterns. PPTA measures spatial correlations consistent with a GWB at \(\sim2\sigma\) significance (false-alarm probability \(\lesssim 0.02\)), following the HD pattern. NANOGrav detects a stochastic GWB with HD correlations at \(3\sigma\) (\(p \approx 10^{-3}\)) via Bayesian analysis and \(3.5-4\sigma\) via frequentist tests.}. This result is consistent with the presence of a substantial population of SMBHB systems in the Universe. The mergers of SMBHB are the main targets for space-borne GW detectors like LISA \citep{amaro2017laser}, Taiji \citep{ruan2020taiji}, and TianQin \citep{mei2021tianqin}. Due to their massive nature, such system can be observed to a much larger distance (100 Gpc, \citep{klein2016science}) than the stellar mass BHs (a few Gpc, \citep{abbott2021tests}). Further more, as the GW from SMBHB mergers are much lower than that from stellar mass BBH mergers, the dispersion effects can be more prominent. Gao et al. \cite{gao2023testing} has simulated GW sources with a signal-to-noise ratio (SNR) of approximately 1000, demonstrating that by utilizing waveform dispersion relations, each of the three detectors alone can constrain the graviton mass to \(10^{-25} \mathrm{eV}/c^2\). If jointly observed, the constraint can be improved to \(10^{-26} \mathrm{eV}/c^2\).  

In this work, we conduct a simulation study to investigate how observations of the cosmic SMBHB population with LISA can help constrain the GW dispersion parameter, thereby placing limits on different gravitational theories. Additionally, we study how joint observations on these sources with electromagnetic counterparts will help to put the constraints on theories beyond GR, including both dispersive and non-dispersive models.

The paper is organised as follows: In Section II, we will use existing SMBHB population data to simulate the GW burst signals of each source. Using the MCMC algorithm, we will provide constraints on the dispersion parameter \(\mathbb{A}_\alpha\) under a given dispersion relation and coefficient \(\alpha\). Finally, we will discuss methods for inferring the value of \(\alpha\) based on observations. In Section III, after simulating the number of sources in the catalog that can be jointly detected, we further calculated the constraints these sources impose on the GW speed and graviton mass. Finally, we will present a summary and discussion. 

\section{Constraining GW Dispersion with SMBHB Populations}

\subsection{SMBHB Populations}

In our study, we employ the SMBHB populations simulated by Klein
et al. \cite{klein2016science} (K16 hereafter for short). In K16, the authors simulate the SMBHB populations by using seed BHs originate from the remnants of Population \MakeUppercase{\romannumeral 3} (pop\MakeUppercase{\romannumeral 3}) stars formed in the low-metallicity environments of the early universe at \( z \approx 15-20 \). By incorporating the delay between massive black hole (MBH) mergers and galaxy mergers, the resulting SMBHB population is referred to as pop\MakeUppercase{\romannumeral 3}; Alternatively, if the seeds arise from the collapse of protogalactic disks, forming more massive seeds of \( 10^5 M_\odot \) (assuming a critical Toomre parameter \( Q_c = 3 \), which gives this model its name), and the merger delay is taken into account, the resulting population is labeled as Q3-d. When the delay is neglected, the resulting population is denoted as Q3-nod. \footnote{It is worth noting that for the light-seed pop\MakeUppercase{\romannumeral 3} population, neglecting the delay leads to variations of less than a factor of two; thus, K16 did not consider this case separately.}

K16 simulated 10 realizations of SMBHB mergers in the Universe within 5 years, which can be equivalently regarded as a catalog of SMBHB mergers over a 50-year period. In this study, we obtain the catalogues of each population from the outputs of \texttt{GW-Universe Toolbox} \citep{yi2022gravitational}.

\begin{figure}
    \centering
    \includegraphics[width=.8\linewidth]{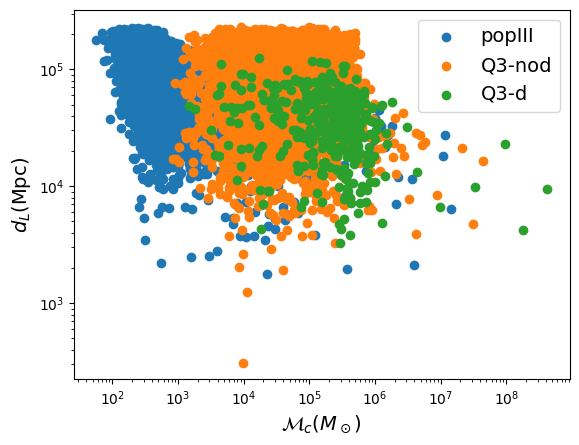}
    \includegraphics[width=.8\linewidth]{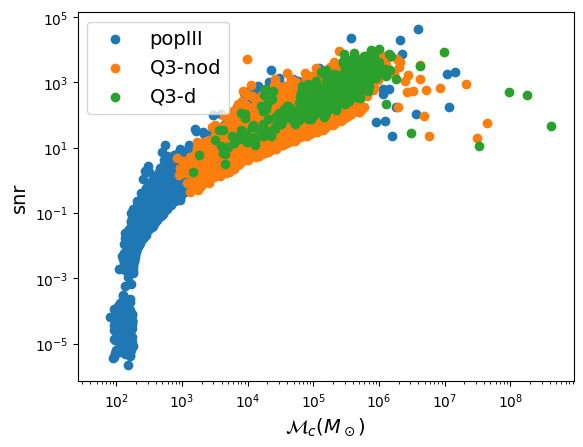}
    \caption{\textbf{The distributions of chirp mass and luminosity distance (upper panel) or SNR (lower panel) for the sources in the three catalogs.}}
    \label{mc_dist}
\end{figure}

The event rate varies among the three populations. In the catalogue corresponding to the pop\MakeUppercase{\romannumeral 3} population, which is the most optimistic, the event rate is 174.7 \(\rm year^{-1}\) containing 8,735 sources; while Q3-d is the most pessimistic, with only 409 sources, with the event rate 8.18 \(\rm year^{-1}\), and Q3-nod represents an intermediate scenario, comprising 6,122 sources and an event rate of 122.44 \(\rm year^{-1}\). However, as shown in upper panel of Figure \ref{mc_dist}, when comparing the distributions of chirp mass and luminosity distance, we observe that while the luminosity distances of the sources do not differ significantly, Q3-d contains sources with systematically larger chirp masses, whereas pop\MakeUppercase{\romannumeral 3} contains smaller ones.  

We employ \texttt{LISAanalysistools} \citep{michael_katz_2024_10930980} and Eq. \ref{phi_mg} to generate waveforms for each SMBHB event in the corresponding population. The waveforms use the frequency-domain \texttt{PhenomHM} model \citep{london2018first, husa2016frequency, khan2016frequency}, which includes higher-order modes beyond the dominant quadrupole. The LISA detector response is modeled with the fast frequency-domain TDI response function described in \cite{marsat2018fourier, marsat2021exploring}, implemented within \texttt{LISAanalysistools}. The signal-to-noise ratio (SNR) for each source is calculated using the default sensitivity matrix for the A, E, and T channels of the Time-Delay Interferometry (TDI) combinations \cite{babak2021lisa}, which accounts for instrumental noise and detector response based on the LISA Consortium Proposal for L3 mission \cite{amaro2017laser}. Taking SNR = 5 as a reference threshold \citep{arun2022new} of detection and SNR = 100 as a conservative threshold, we find that population model pop\MakeUppercase{\romannumeral 3} results in fewest number of detection. The total number of detection corresponding to each population models are sumarized in Table \ref{3cata}.

\begin{table}
\caption{\textbf{Table for Characteristics and number of three catalogs}: The second column describes the characteristics of each source catalog, the third column lists the total number of sources across 10 catalogs, each representing merger events within a 5-year period, the last two columns show the number of sources with a SNR greater than 5 and 100}
\setlength{\tabcolsep}{24pt}
\begin{tabular}{ccccc}
\hline
name                                                               & characteristics                                                          & \begin{tabular}[c]{@{}c@{}}total \\ number\end{tabular} & \begin{tabular}[c]{@{}c@{}}number\\  of snr\textgreater{}5\end{tabular} & \begin{tabular}[c]{@{}c@{}}number\\  of snr\textgreater{}100\end{tabular} \\ \hline
pop\MakeUppercase{\romannumeral 3} & \begin{tabular}[c]{@{}c@{}}light MBH seeds\\  with delay\end{tabular}    & 8735                                                    & 567                                                                       & 123       \\
Q3-d                                                               & \begin{tabular}[c]{@{}c@{}}heavy MBH seeds\\  with delay\end{tabular}    & 409                                                     & 407                                                                       & 348     \\
Q3-nod                                                             & \begin{tabular}[c]{@{}c@{}}heavy MBH seeds\\  without delay\end{tabular} & 6122                                                    & 5917                                                                      & 2022      \\ \hline
\end{tabular}
\label{3cata}
\end{table}

\subsection{Constraints on Dispersion Coefficients with given \(\alpha\) with single source}

Using the generated waveforms, we use \texttt{LISAanalysistools} to simulate LISA's response to SMBHB mergers and compute the likelihood \(\mathcal{L}\) as function of GW parameters:
\begin{equation}
    \begin{aligned}
        \mathcal{L}&\propto-1/2\langle d-h|d-h\rangle\\
        &=-1/2(\langle d|d\rangle\langle h|h\rangle-2\langle d|h\rangle)\,,
    \end{aligned}
\end{equation}
where \(d\) is the measured waveform obtained
from the injected source parameters, while \(h\) is the parametric GW template
model. To lower the required memory and increase the speed of the computation, we adopt the heterodyned method \citep{cornish2021heterodyned, zackay2018relative}. With the simulated LISA response data, Bayesian inference can then be applied to constrain the model parameters. Given the likelihood, we employ the \texttt{Eryn} \citep{michael_katz_2023_7705496} package to perform parallel-tempered MCMC sampling. Due to the relatively simple target distribution and prior knowledge of the parameter ranges, we use 3 temperatures and 32 walkers to sample a 12-dimensional parameter space. The number of steps required for convergence varies between 2000 and 20000, corresponding to computation times from approximately 20 minutes to 4 hours per source. The simulated data \(d\) include both the gravitational-wave signal and random noise realizations. The noise is generated using the A, E, and T channel sensitivity matrix implemented in the \texttt{AET1SensitivityMatrix} module of \texttt{LISAanalysistools}.

Next, we will demonstrate the fitting results for a single source, using the source with the highest SNR as an example. The parameters of this source are presented in Table \ref{value1570}.

\begin{table}
\centering
\caption{\textbf{Table of the parameter values of the source with the highest SNR}: From top to bottom, the parameters are: redshift, primary star mass, secondary star mass, chirp mass, dimensionless spin of \(m_1\), dimensionless spin of \(m_2\), luminosity distance, ecliptic longitude, ecliptic latitude, inclination of the binary, polarization angle, and SNR.}
\setlength{\tabcolsep}{90pt}
\begin{tabular}{cc}
\hline
parameters & value                  \\ \hline
$z$        & 0.383                  \\
$m_1/M_\odot$       & $6.36\times10^{6}$              \\
$m_2/M_\odot$       & $3.32\times10^{6}$              \\
$\mathcal{M}_c/M_\odot$       & $3.96\times10^{6}$       \\
$\chi_1$     & 0.96                 \\
$\chi_2$     & 0.90                \\
$d_L$/m     & $6.58\times10^{25}$  \\
$\lambda$/rad   & 4.98                 \\
$\beta$/rad     & -0.0017 \\
$i$/rad      & 0.3927                 \\
$\psi$/rad      & 1.1779                 \\
SNR      & 44175      \\ \hline
\end{tabular}
\label{value1570}
\end{table}
 
\begin{figure}
    \centering
    \includegraphics[width=.8\linewidth]{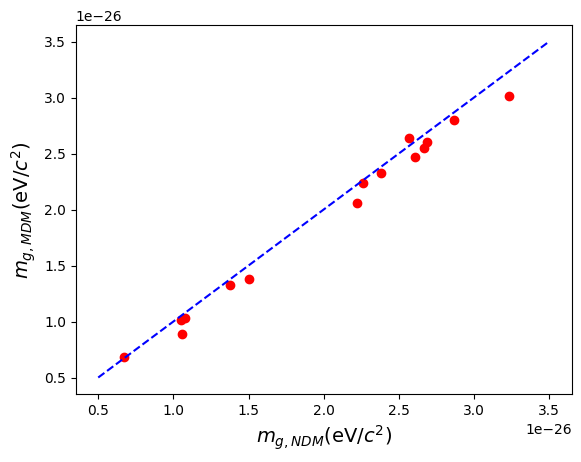}
    \caption{\textbf{The plot shows the upper limits on the graviton mass constrained using two different methods}: The blue dashed line represents the reference line where the two values are equal. The horizontal axis corresponds to the upper limit on the graviton mass constrained using non-dispersive waveforms, {\it i.e.,} Non-dispersion mass (NDM), and the vertical axis corresponds to the upper limit on the graviton mass that can be detected, {\it i.e.,} minimum detectable mass (MDM).}
    \label{2mg}
\end{figure}

The bounds of the dispersion coefficients are determined by progressively increasing the absolute value of $|\mathbb{A}_\alpha|$ in Eq. \ref{phi_mg} and examining the resulting posterior distributions. When the 90\% credible interval of the posterior excludes zero, we consider the dispersion coefficient to be nonzero. We refer the bounds found with this method "the minimum detectable bounds (MDB)". The MDB method need to loop with increasing $|\mathbb{A}_\alpha|$ and apply Bayesian analysis in each round of loop and therefore is computational expensive. Another bounds of the dispersion coefficients can be defined as the bounds corresponds to certain quantiles (we apply 90\%) in the posterior distribution of $\mathbb{A}_\alpha$ obtained against waveform with zero dispersion. We refer this bounds as non-dispersion bounds (NDB). In Figure \ref{2mg}, we show the plot of the graviton mass constraints for 15 sources from the catalog using the two methods mentioned above. In the figure, we found that these two values are very close to each other. Given that MCMC sampling inherently involves some uncertainty, using the second value to represent the first one is a reasonable approach. Therefore, in the following sections, we adopt the NDB method for all analyses, since it is equivalent to the MDB method.

As discussed above, when \(\alpha=0\) and \(\mathbb{A}_0>1\), the dispersion coefficient \(\mathbb{A}_0\) corresponds to \(m_g^2\). Instead of directly inferring \(m_g\), which is in large degeneracy with the chirp mass and luminosity distance, we do inference on the \(\beta_0\) as defined in Eq. \ref{phi_mg} and the equation following it, and convert the posterior of \(\beta_0\) to \(m_g\). Moreover, to avoid boundary effects impacting the posterior distribution of parameters, we retained the \(\beta_0 < 0\) region in the prior for \(\beta_0\). This part is discarded only when transforming \(\beta_0\) into \(m_g\) at the final stage.

For the other dispersion parameters, similar to the case of the graviton mass, we fit the parameter \(\zeta\) and subsequently convert it into the desired \(\mathbb{A}_\alpha\). Unlike the graviton mass, these dispersion parameters are not restricted in sign, allowing us to retain all values from the MCMC chains.

\begin{figure*}
    \centering
    \includegraphics[width=\linewidth]{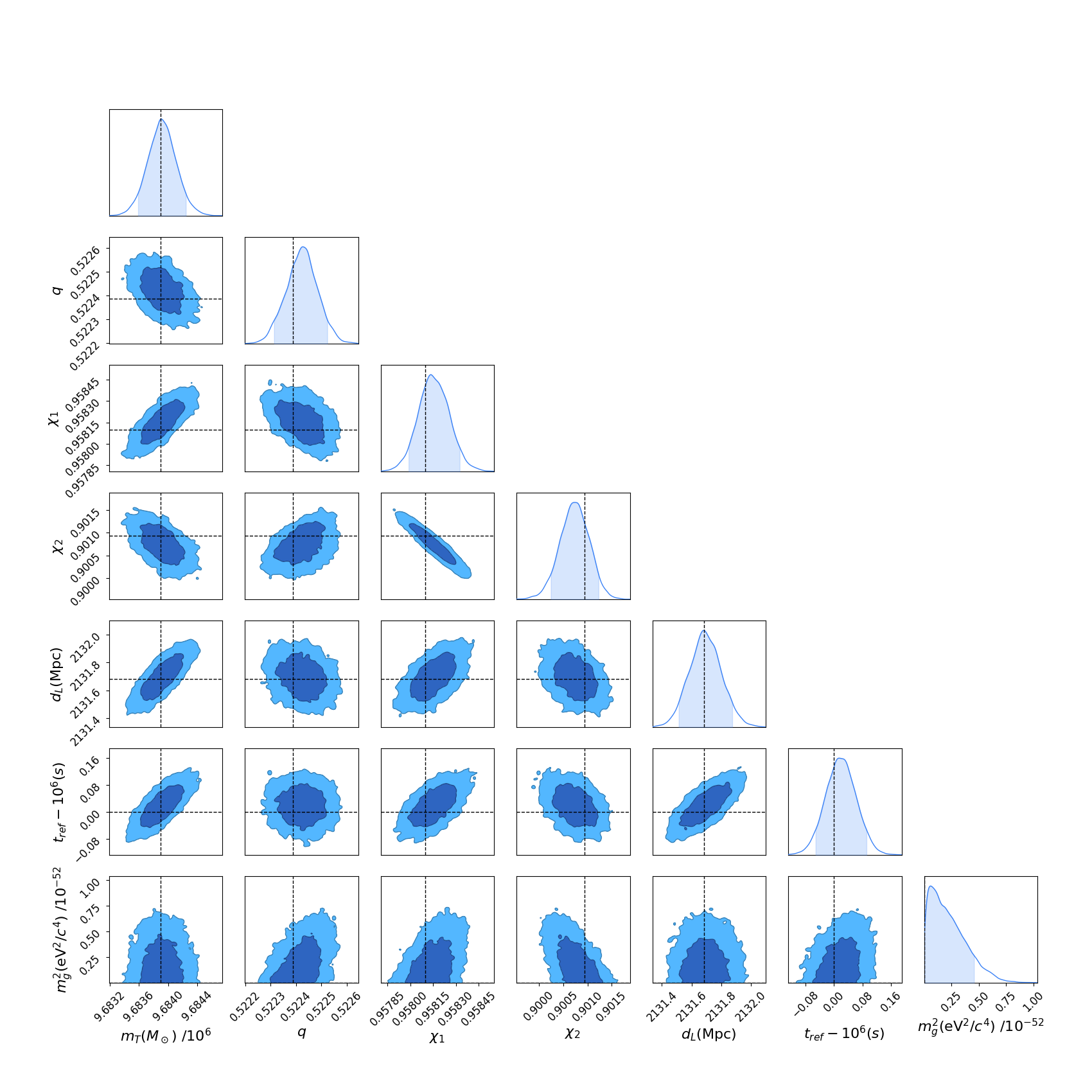}
    \caption{\textbf{Corner figure of the posterior distributions for the parameters of the example source}: The parameters are in order: total mass $m_{\rm T}$ in unit of \(M_\odot\), mass ratio $q\equiv m_2/m_1$, dimensionless spin $\chi_1$ of the primary BH, dimensionless spin of $\chi_2$ of the secondary BH, luminosity distance $d_{\rm L}$ in unit of Mpc, the reference time $t_{\rm ref}$ corresponding to the frequency at which the signal’s energy output is maximal in unit of s and the square of the graviton mass $m^2_g$ in unit of (\(\mathrm{eV}/c^2)^2\).}
    \label{cor1570}
\end{figure*}

Finally, the posterior distributions of the parameters for the source with the highest SNR are shown in Figure \ref{cor1570}. In this analysis, the injected graviton mass is set to zero. The variable \(\beta_0\) has already been transformed into 
\(m_g^2\). The results of all dispersion parameter constraints, including the graviton mass, are presented in Table \ref{result1570}.

\subsection{Joint Analysis on $\mathbb{A}_\alpha$ with multiple events in a population}

\begin{figure}
    \centering
    \includegraphics[width=.6\linewidth]{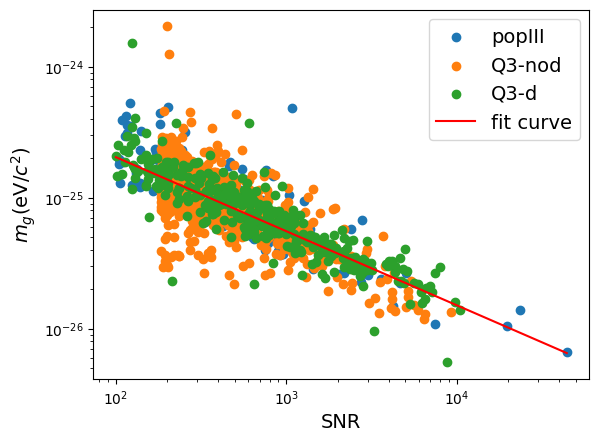}
    \includegraphics[width=.6\linewidth]{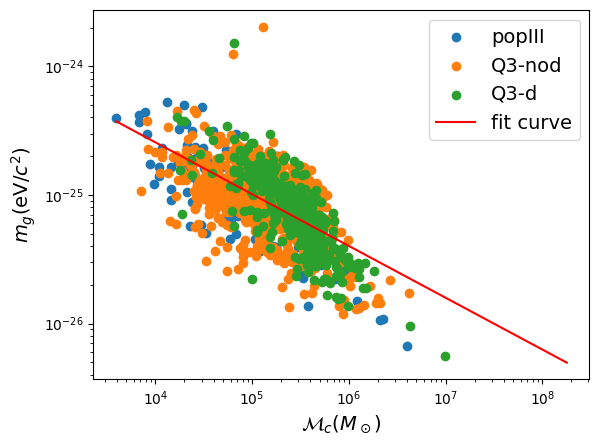}
    \includegraphics[width=.6\linewidth]{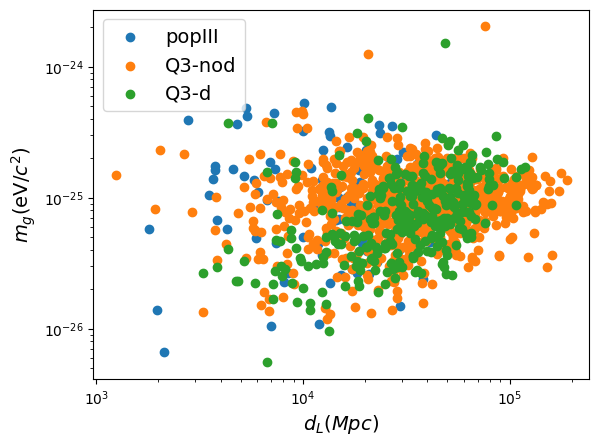}
    \caption{\textbf{The distributions of the upper limit on the graviton mass and SNR (upper panel), chirp mass (middle panel) or luminosity distance (lower panel) for the sources in the three catalogs.}}
    \label{snr_mg}
\end{figure}

The above methods can be applied to all sources in the three catalogs, and a joint constraining with multiple sources on $\mathbb{A}_\alpha$ can be naturally expected to be better than that from a single source. After computing the upper limit on the graviton mass for all sources with SNR greater than 100, we plotted the upper limit of the graviton mass against the SNR, chirp mass, and luminosity distance, as shown in Figure \ref{snr_mg}.

In the figure, we observe that the SNR and the upper limit on the graviton mass from individual source exhibit a strong linear relationship in logarithmic space. The slope is approximately -0.5, indicating that increasing the SNR by two orders of magnitude improves the upper limit on the graviton mass by one order of magnitude.  

Similarly, there is a certain degree of linear correlation between the chirp mass and the upper limit on the graviton mass in logarithmic space. As shown in Figure \ref{snr_mg}, the SNR and chirp mass also exhibit a linear relationship in logarithmic space. The slope is approximately -0.4, implying that achieving an equivalent improvement in the graviton mass constraint requires a greater increase in the chirp mass.   

Regarding the relationship between luminosity distance and the upper limit on the graviton mass, no significant correlation is observed in the figure. The detected sources are primarily distributed at luminosity distances of around \(10^4 - 10^5\) Mpc, which is consistent with the initial distribution of the three catalogs. Other parameters, similar to luminosity distance, do not show a significant correlation with the upper limit on the graviton mass. Their distributions also do not exhibit noticeable differences from the initial distributions. Therefore, we do not elaborate further on these aspects. 

Theoretically, the variation $\Delta\tilde{h}$ in the signal that can be constrained by the same GW detector remains essentially constant, and can be simplified as
\begin{equation}
\begin{aligned}
\Delta\tilde{h}&=\frac{A(\tilde{t})}{\sqrt{\dot{f}(\tilde{t})}}e^{i\Psi_{\rm GR}+i\delta\Psi(\tilde{f})}-\frac{A(\tilde{t})}{\sqrt{\dot{f}(\tilde{t})}}e^{i\Psi_{\rm GR}}\\
&=\frac{A(\tilde{t})}{\sqrt{\dot{f}(\tilde{t})}}e^{i\Psi_{\rm GR}}\cdot[e^{i\delta\Psi(\tilde{f})}-1]\\
&\sim A(\tilde{t})\delta\Psi(\tilde{f}),.
\end{aligned}
\end{equation}
From Eq. \ref{phi_zeta},\ref{zeta},\ref{D_alpha}, it follows that
\begin{equation}
\delta\Psi(\tilde{f})\sim d_L\mathbb{A}_\alpha,,
\end{equation}
where we approximate $D_\alpha/(1+Z)^{1-\alpha}$ by $d_L$. This result is independent of the chirp mass $\mathcal{M}$. The GW amplitude is proportional to
\begin{equation}
A(\tilde{f})\propto{\mathcal{M}^{5/6}\over d_L},.
\end{equation}
Combining the above, we obtain
\begin{equation}
\Delta\tilde{h}\sim \mathbb{A}_\alpha\mathcal{M}^{5/6},.
\end{equation}

Assuming $\Delta\tilde{h}$ is approximately constant, the constraint on $\mathbb{A}_\alpha$ is given by
\begin{equation}
\mathbb{A}_\alpha\sim\mathcal{M}^{-5/6},,
\end{equation}
and, in the special case of $\alpha=0$, $m_g=\sqrt{\mathbb{A}_0}\sim\mathcal{M}^{-5/12}\approx\mathcal{M}^{-0.4}$. This is consistent with the results in Fig. \ref{snr_mg}, and likewise explains why the constraint on $m_g$ is independent of the luminosity distance.

For the SNR,
\begin{equation}
\Delta\tilde{h}\sim SNR\cdot\delta\Psi(\tilde{f})=SNR\cdot\mathbb{A}_\alpha,,
\end{equation}
we have $\mathbb{A}_\alpha\sim SNR^{-1}$ and $m_g=\sqrt{\mathbb{A}_0}\propto SNR^{-0.5}$, which are also consistent with the results in Fig. \ref{snr_mg}.

Based on the above derivation, the distribution of $\mathbb{A}_\alpha$ for different values of $\alpha$ follows the same trend as in the $\alpha=0$ case. We also computed the distributions of other $\mathbb{A}_\alpha$ parameters for the sources in the three catalogs. These distributions exhibit similar trends to that of the graviton mass $m_g$, and thus we do not elaborate further here.

\begin{figure}
    \centering
    \includegraphics[width=.8\linewidth]{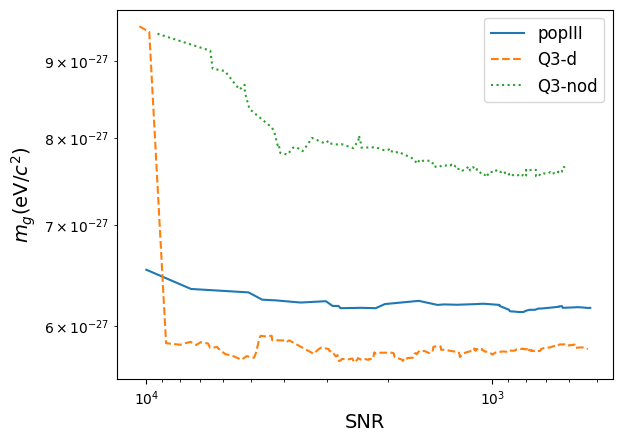}
    \caption{\textbf{The upper limit on the graviton mass constrained by space-based detectors (assuming LISA's noise) under different SNR thresholds.}}
    \label{mg_N}
\end{figure}

Since the propagation effects of individual source are independent, we can obtain the joint probability distribution of multiple sources by multiplying their posterior probabilities \citep{abbott2016tests}. We plotted the results of the joint constraint on the graviton mass under different SNR thresholds, as shown in Figure \ref{mg_N}.

\begin{figure*}
    \centering
    \includegraphics[width=.4\linewidth]{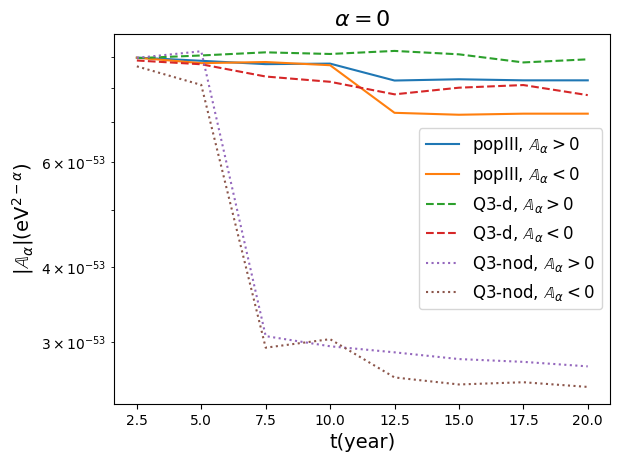}
    \includegraphics[width=.4\linewidth]{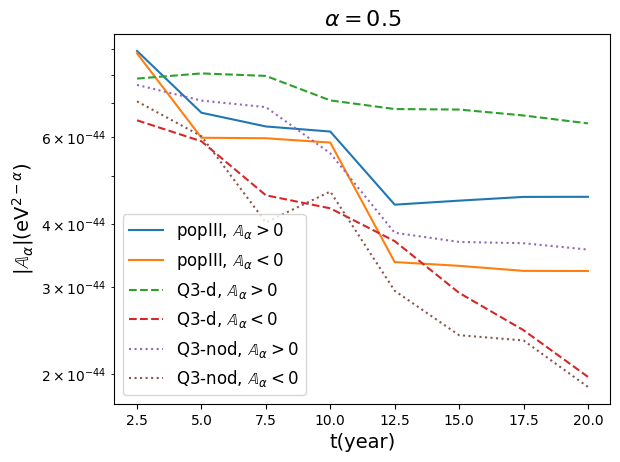}
    \includegraphics[width=.4\linewidth]{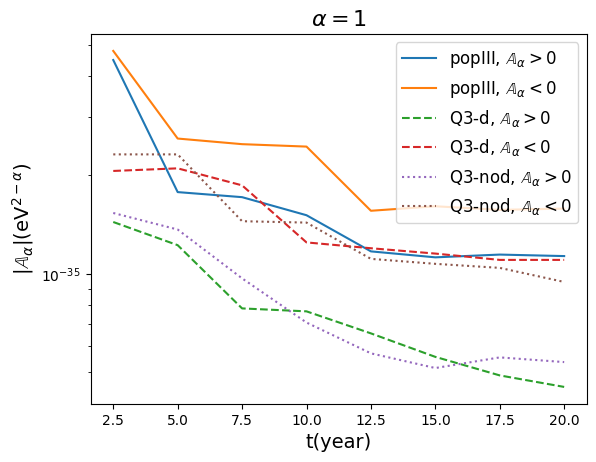}
    \includegraphics[width=.4\linewidth]{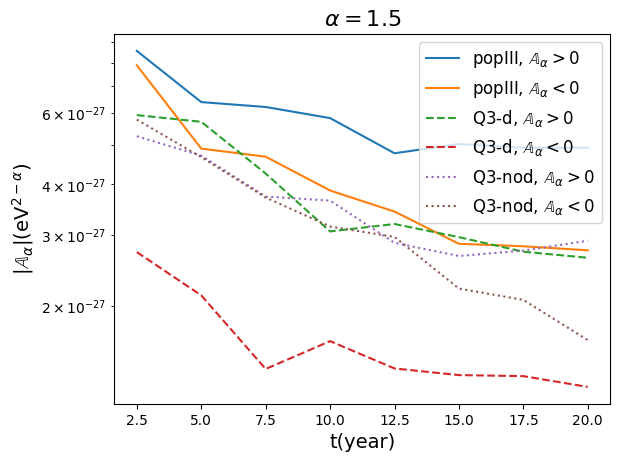}
    \includegraphics[width=.4\linewidth]{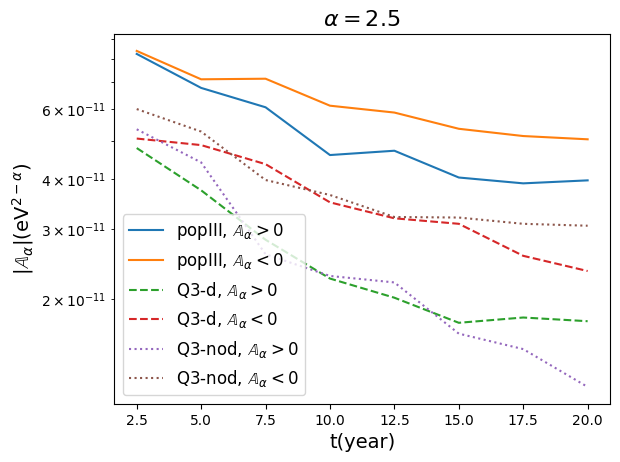}
    \includegraphics[width=.4\linewidth]{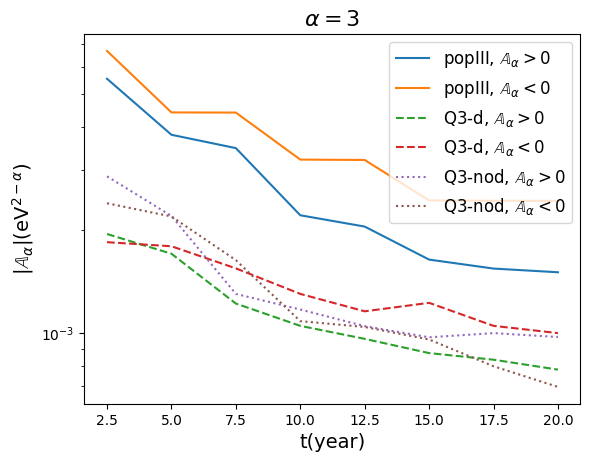}
    \includegraphics[width=.4\linewidth]{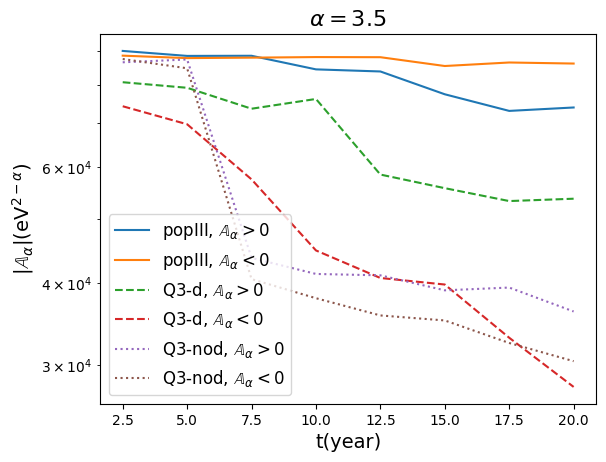}
    \includegraphics[width=.4\linewidth]{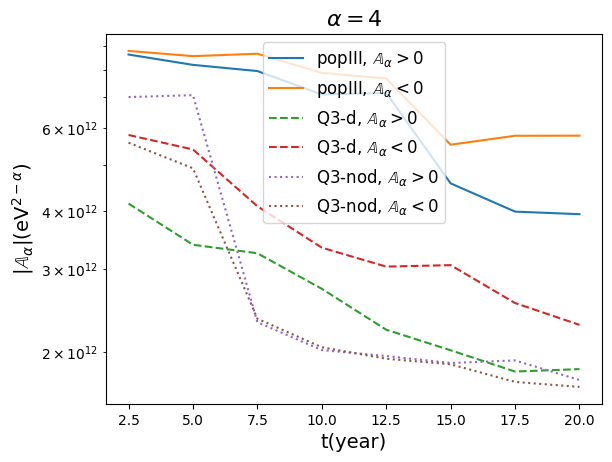}
    \caption{\textbf{The relationship between the upper limit on the graviton mass constrained by LISA and the mission duration}: Source catalogs are taken from K16, which provides 10 independent 5-year SMBHB source lists. To simulate a 2.5-year observation, each catalog is randomly split into two subsets. For mission durations longer than 5 years, additional catalogs are randomly added to the original list.}
    \label{A_alpha_time}
\end{figure*}

Analysis of the figure reveals that for catalogs such as pop\MakeUppercase{\romannumeral 3} and Q3-d, a small number of exceptionally constraining sources dominate the results, causing the constraints to plateau once these sources are detected; further decreasing the SNR threshold yields minimal improvement. Conversely, for more uniformly distributed catalogs like Q3-nod, the constraints improve steadily and stabilize only when the SNR threshold approaches 1000. These findings indicate that the joint constraints on $\mathbb{A}_\alpha$ are effectively saturated with sources having $\mathrm{SNR} > 1000$, and the addition of lower-SNR sources contributes marginally.

We also note there is fluctuation around this average trend. It is because although the constraint on \(m_g\) is monotonously tighten with more sources included, the upper limit on its positive value is not unnecessarily less correspondingly. 

Using the same method, we have calculated the 90\%-credible bounds for \(\mathbb{A}_\alpha\) in eight different cases: \(\alpha = 0, 0.5, 1, 1.5, 2.5, 3.5, 4\), and presented its evolution over time in Figure \ref{A_alpha_time}. According to the conclusions from the Figure \ref{mg_N}, we only considered the sources with the SNR larger than 1000. 

Similar to the conclusions drawn for the graviton mass, certain exceptionally well-constrained sources play a decisive role. In general, there will always be a few high-SNR sources contributing to the final result in each time interval. Therefore, as time progresses, the constraint results will always improve, but the degree of improvement depends on whether exceptionally well-constrained sources are detected.

We compare our results with those from GWTC-3 by selecting the constraints obtained from 5 and 10 years of LISA observations and presenting them alongside the GWTC-3 results in Table \ref{result}. We find that for LISA, smaller values of \(\alpha\) generally lead to stronger constraints compared to GWTC-3. However, when \(\alpha > 2\), the constraints become significantly weaker than those from GWTC-3. This is understandable, as seen from Eq. \ref{phi_zeta}, where \(\delta\Psi\) is proportional to \(f^{\alpha-1}\) (or \(\ln{f}\) when \(\alpha=1\)). Since LISA operates at lower frequencies, it is naturally more sensitive to smaller values of \(\alpha\).

\begin{sidewaystable*}
\centering
\caption{\textbf{Table showing the 90\%-credible bounds of LISA's constraints on \(m_g\) and \(\mathbb{A}_\alpha\) for 5 or 10 years of observation in the case of three different SMBHB catalogs}: The table also includes the results from GWTC-3 \citep{abbott2021tests}. Based on the expected observation duration of LISA and the observation time of GWTC-3, we provide the constraint results for 5 and 10 years of LISA observation. For \(\mathbb{A}_\alpha\), the unit is \(\mathrm{eV}^{2-\alpha}\), but due to space limitations, we have not listed this in the table. Additionally, since there is a significant difference in the magnitude between the GWTC-3 and LISA constraint results, we separately display the order of magnitude for the GWTC-3 results in the last row.}
\setlength{\tabcolsep}{7pt}
\begin{tabular}{cccccccccccccccccc}
\hline
\multirow{2}{*}{} & \multirow{2}{*}{\begin{tabular}[c]{@{}c@{}}\(m_g(\mathrm{eV}/c^2)\)\\ (\(10^{-27}\)) \end{tabular}} & \multicolumn{2}{c}{\begin{tabular}[c]{@{}c@{}}\(\mathbb{A}_{0}\)\\ (\(10^{-53}\)) \end{tabular}} & \multicolumn{2}{c}{\begin{tabular}[c]{@{}c@{}}\(\mathbb{A}_{0.5}\)\\ (\(10^{-44}\))\end{tabular}} & \multicolumn{2}{c}{\begin{tabular}[c]{@{}c@{}}\(\mathbb{A}_{1}\)\\ (\(10^{-35}\))\end{tabular}} & \multicolumn{2}{c}{\begin{tabular}[c]{@{}c@{}}\(\mathbb{A}_{1.5}\)\\ (\(10^{-27}\))\end{tabular}} & \multicolumn{2}{c}{\begin{tabular}[c]{@{}c@{}}\(\mathbb{A}_{2.5}\)\\ (\(10^{-11}\))\end{tabular}} & \multicolumn{2}{c}{\begin{tabular}[c]{@{}c@{}}\(\mathbb{A}_{3}\)\\ (\(10^{-3}\))\end{tabular}} & \multicolumn{2}{c}{\begin{tabular}[c]{@{}c@{}}\(\mathbb{A}_{3.5}\)\\ (\(10^{4}\))\end{tabular}} & \multicolumn{2}{c}{\begin{tabular}[c]{@{}c@{}}\(\mathbb{A}_{4}\)\\ (\(10^{12}\))\end{tabular}} \\ \cline{3-18}
& & + & - & + & - & + & - & + & - & + & - & + & - & + & - & + & - \\ \hline
\begin{tabular}[c]{@{}c@{}}pop\MakeUppercase{\romannumeral 3}\\ (5 years)\end{tabular} & 9.50 & 9.02 & 8.98 & 9.00 & 8.92 & 8.12 & 8.87 & 9.23 & 8.59 & 8.59 & 9.20 & 5.07 & 9.03 & 9.01 & 8.93 & 8.89 & 9.01 \\ 
\begin{tabular}[c]{@{}c@{}}pop\MakeUppercase{\romannumeral 3}\\ (10 years)\end{tabular} & 6.39 & 4.08 & 5.24 & 2.35 & 2.39 & 1.01 & 0.98 & 3.52 & 3.20 & 4.56 & 5.43 & 2.23 & 3.26 & 8.87 & 8.62 & 8.40 & 8.89 \\ 
\begin{tabular}[c]{@{}c@{}}Q3-d\\ (5 years)\end{tabular} & 9.33 & 8.71 & 8.14 & 5.35 & 4.23 & 1.09 & 1.27 & 3.76 & 2.61 & 4.13 & 5.29 & 1.59 & 2.82 & 8.29 & 8.39 & 7.62 & 7.22 \\ 
\begin{tabular}[c]{@{}c@{}}Q3-d\\ (10 years)\end{tabular} & 5.89 & 3.47 & 3.01 & 2.48 & 2.05 & 0.74 & 0.98 & 3.08 & 2.12 & 3.29 & 3.69 & 1.05 & 2.25 & 7.80 & 8.25 & 7.56 & 5.20 \\ 
\begin{tabular}[c]{@{}c@{}}Q3-nod\\ (5 years)\end{tabular} & 9.05 & 8.19 & 8.40 & 4.47 & 4.38 & 1.38 & 1.69 & 4.53 & 5.51 & 7.12 & 6.60 & 2.94 & 3.62 & 8.70 & 8.95 & 7.49 & 7.98 \\ 
\begin{tabular}[c]{@{}c@{}}Q3-nod\\ (10 years)\end{tabular} & 8.88 & 7.89 & 7.54 & 3.55 & 2.94 & 0.98 & 1.05 & 3.18 & 3.40 & 5.28 & 5.22 & 1.64 & 2.48 & 7.55 & 8.43 & 6.70 & 6.46 \\ \hline
\multirow{2}{*}{GWTC-3} & 1.27 & 0.89 & 1.88 & 0.19 & 0.51 & 0.32 & 1.16 & 0.93 & 3.69 & 2.95 & 1.16 & 2.33 & 0.66 & 1.16 & 0.45 & 0.74 & 0.30 \\
& (\(10^{-23}\)) & \multicolumn{2}{c}{(\(10^{-45}\))} & \multicolumn{2}{c}{(\(10^{-38}\))} & \multicolumn{2}{c}{(\(10^{-32}\))} & \multicolumn{2}{c}{(\(10^{-26}\))} & \multicolumn{2}{c}{(\(10^{-14}\))} & \multicolumn{2}{c}{(\(10^{-8}\))} & \multicolumn{2}{c}{(\(10^{-2}\))} & \multicolumn{2}{c}{(\(10^{4}\))} \\ \hline
\end{tabular}
\label{result}
\end{sidewaystable*}

\begin{figure}
    \centering
    \includegraphics[width=.55\linewidth]{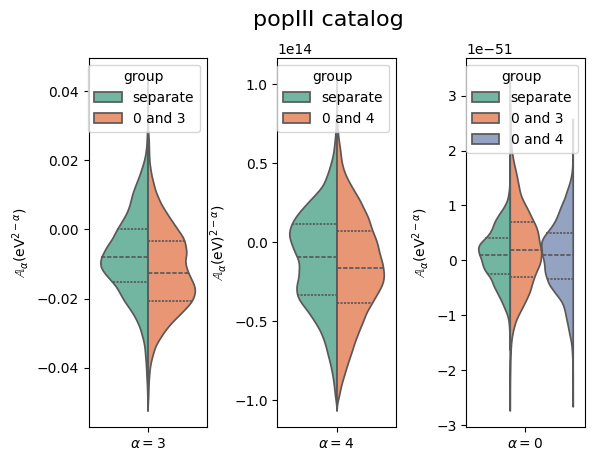}
    \includegraphics[width=.55\linewidth]{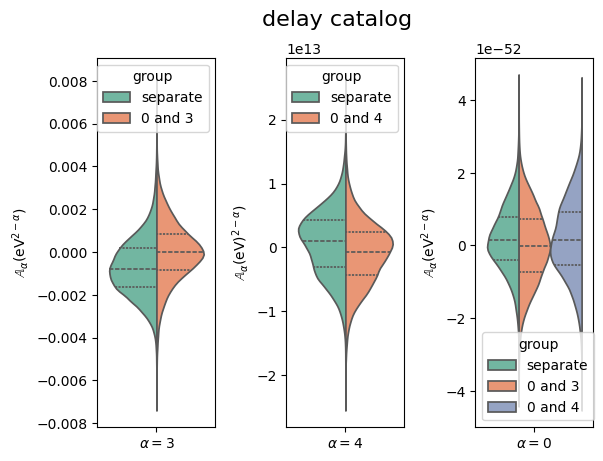}
    \includegraphics[width=.55\linewidth]{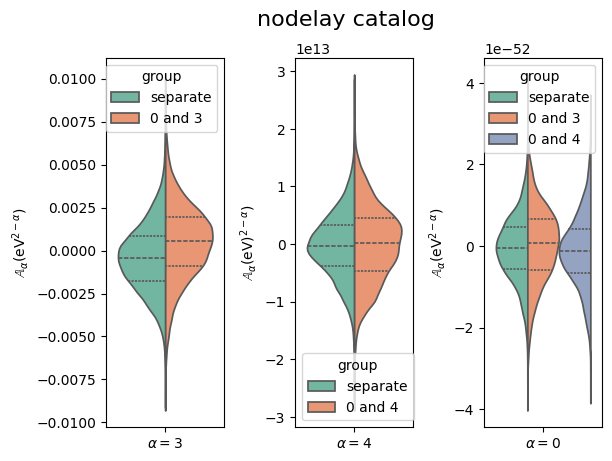}
    \caption{\textbf{Violin plot of \(\mathbb{A}_\alpha\) distribution when fitting one or multiple \(\alpha\) values}: We plotted the results separately for \(\alpha = 0, 3, 4\), labeled as 'separate', and for cases where \(\alpha = 0,3\) or \(\alpha = 0,4\) were fitted simultaneously, labeled as '0 and 3' and '0 and 4', respectively. All results correspond to a 5-year observation period.}
    \label{violin}
\end{figure}

In the theories listed in Table \ref{theory}, DSR, ED, and NCG also allow for the simultaneous existence of \(\alpha=0\) and \(\alpha=3,4\). Therefore, we have also conducted simulations for these cases, with the results shown in Figure \ref{violin}. In the figure, we can see that when fitting two \(\alpha\) values simultaneously, the final results do not change significantly. Therefore, regardless of how many \(\alpha\) values a theoretical model predicts, it is reasonable to sample them separately.

\subsection{Inference of Dispersion Parameters with Unknown \(\alpha\)}

In the previous section, we constrained the corresponding coefficient \(\mathbb{A}_\alpha\) using Bayesian inference under the assumption that \(\alpha\) is known. However, in practice, the true value of \(\alpha\) is unknown and is one of the parameters we aim to determine. Since \(\alpha\) takes discrete values, it is not suitable to treat \(\alpha\) as a free parameter in the fitting process. Therefore, we assume that the \(\alpha\) values considered in the previous section already cover all possible cases, and our task is to identify the most probable model among these eight options. 

The Bayes evidence is one of the most commonly used metrics to quantify the support for one model. We can use the ratio of the evidences of two models, also known as the Bayes factor, to assess the relative preference between the two models.

Accordingly, using the source with parameters in Table \ref{value1570} as an example, we simulate the corresponding dispersion-induced waveform with injected \(\mathbb{A}_\alpha\) corresponds to its upper limit from the previous section. Here, we employed the Thermodynamic Integration method provided by \texttt{Eryn} to estimate the evidence. According to the precision requirements of thermodynamic integration \cite{lartillot2006computing} and considering computational resources, we used 20 temperatures and 24 walkers to fit the waveform with eight different models and to compute the Bayes factor for each, as shown in Table \ref{evidence}.

\begin{table}
\caption{\textbf{The logarithmic Bayesian factor obtained from full Bayesian inference using different dispersion relations}: Each row represents the logarithm of the Bayes factor, where the Bayes factor is calculated by comparing the model with a given dispersion relation corresponding to the row’s \(\alpha\) value against models with different \(\alpha\) values.}
\setlength{\tabcolsep}{11pt}
\begin{tabular}{ccccccccc}
\hline
\diagbox{$\alpha_{\rm real}$}{$\alpha_{\rm model}$} & 0      & 0.5         & 1      & 1.5    & 2.5         & 3      & 3.5         & 4      \\ \hline
0     & \textbf{0.00} & -0.08 & -0.45 & -0.43 & -0.69 & -0.83 & -0.88 & -0.90 \\
0.5   & -0.01 & \textbf{0.00} & -0.02 & -0.10 & -0.42 & -0.50 & -0.57 & -0.70 \\
1     & -0.58 & -0.00 & \textbf{0.00} & -0.18 & -0.72 & -0.65 & -0.78 & -1.27 \\
1.5   & -1.17 & -0.13 & -0.56 & \textbf{0.00} & -1.44 & -1.60 & -2.50 & -0.77 \\
2.5   & -1.74 & -1.15 & -1.06 & -0.23 & \textbf{0.00} & -0.07 & -0.22 & -0.27 \\
3     & -3.40 & -2.48 & -1.68 & -0.64 & -0.14 & \textbf{0.00} & -0.58 & -0.55 \\
3.5   & -4.46 & -3.28 & -2.28 & -1.28 & -0.49 & -0.09 & \textbf{0.00} & -0.14 \\
4     & -2.53 & -2.00 & -1.59 & -0.85 & -0.26 & -0.16 & -0.07 & \textbf{0.00} \\ \hline
\end{tabular}
\label{evidence}
\end{table}

From the table, we can see that fitting with the true value of \(\alpha\) generally yields the highest evidence. However, compared with neighboring values of \(\alpha\), the Bayes factor is often less than 1, and in some cases even below 0.1. Therefore, although this method allows for a preliminary estimation of \(\alpha\), it still cannot uniquely determine its true value. In this section, the injected value of \(\mathbb{A}_\alpha\) corresponds to the minimum value that can be identified. If a value larger than this critical threshold were detected, the Bayes factor would exhibit a clearer distinction.

This implies that the constraint on $\alpha$ for this source is actually beyond the 0–4 range discussed here. However, due to the discrete nature of $\alpha$, we cannot provide a confidence interval for $\alpha$ in the same way as for the dispersion coefficient. Machine learning offers an alternative approach, as it does not require explicit analytical relations and can therefore be applied to cases like $\alpha$ that are discrete. Nevertheless, at present, machine learning results still require verification by traditional Bayesian methods to establish their validity, as results obtained solely through machine learning are not widely accepted. Therefore, further research is needed to obtain a constraint on $\alpha$ that is accepted by the broader community.

\section{Multi-Messenger Constraints on GW Propagation}

In addition to GWs, the accretion flows in SMBHB systems will also produce electromagnetic emission spanning from radio to X-ray bands. Single supermassive black holes accreting at moderately high rates are thought to consist of a geometrically thin, optically thick Novikov–Thorne disk \citep{novikov1973astrophysics}, and an optically thin hot corona that emits a power-law spectrum at higher energies.

However, the radiation from a SMBHB during the pre-merger inspiral phase is more complex, primarily originating from the circumbinary disk \citep{noble2012circumbinary,shi2012three} and the mini-disks formed within the cavity carved out at the binary’s center of mass \citep{shi2015three}. As a result, we can distinguish electromagnetic signals from binary black holes based on their multi-peaked spectra, softer UV emission compared to single black holes, and characteristic quasi-periodic variability in their light curves \citep{gutierrez2022electromagnetic}.

After the black holes merge, a larger central cavity is formed, leading to a reduction in radiative efficiency. As gas flows in, a brightening occurs both before and after the merger. Dong-Páez et al. \cite{dong2023multi} suggests that the radiation at this stage can be modeled using a single supermassive black hole framework, with the luminosity peaking at the Eddington limit.
\begin{equation}  
    L = L_{\rm edd} = \frac{4\pi GM_{\bullet}m_p c}{\sigma_T}\,,  
\end{equation}  
where \( M_{\bullet} \) is the total mass of the system, \( m_p \) is the proton mass, and \( \sigma_T \) is the Thomson scattering cross-section.

We assume that the detection of a brightening, combined with the spectral and temporal characteristics observed prior to the merger, is sufficient to identify the event as originating from a supermassive black hole binary merger. Dong-Páez et al. \cite{dong2023multi} provides three conditions for determining whether detectors in different bands can detect such burst signals
\begin{itemize}
    \item The flux of radiation emitted during the merger of black hole is greater than the radiation flux from its host galaxy; \\
    \item The flux of radiation emitted during the merger of black hole must exceed the sensitivity threshold of the instrument; \\
    \item The flux variation must exceed a factor of 2, or show a transition from non-detection to detection or vice versa.
\end{itemize}

In their simulations, it was found that merger signals in the UV band are difficult to observe because they are typically weaker than those of the host galaxy. Transients are also found to be weaker for radio-observable mergers, making detections primarily concentrated in the X-ray band. In fact, there are approximately 4–20\% of the sources expected to have detectable electromagnetic counterparts in the X-ray band. 

In the presence of interstellar medium surrounding the black hole, we must account for its absorption of X-rays. The rest-frame attenuated luminosity can be calculated as:  
\begin{equation}  
    L_\mathrm{X,abs} = \int^{\nu_\mathrm{max}}_{\nu_\mathrm{min}} L_\nu e^{-\sigma_\mathrm{X}(\nu) N_\mathrm{H}} \, {\rm d}\nu \, ,  
\end{equation}  
where \( N_\mathrm{H} \) is the hydrogen column density, taken as \( N_\mathrm{H} = 10^{22} \mathrm{cm}^{-2} \) \citep{gutierrez2022electromagnetic}. The term \( \sigma_\mathrm{X}(\nu) \) represents the X-ray cross sections, calculated using the polynomial fits from  Morrison \& McCammon \cite{morrison1983interstellar}. Assuming a uniform spectral distribution, we perform the integration and obtain a photon transmission efficiency of about 83.8\%. Thus, the flux ultimately reaching the Earth is:
\begin{equation}
    F = \frac{83.8\%L}{4\pi d_L^2}\,,
\end{equation}

With the above method, we can determine the energy reaching Earth at the time of the outburst for each source in the catalog. By the time LISA is operational, the major X-ray observatories will include the Advanced X-ray Imaging Satellite (AXIS) \citep{mushotzky2018axis}, the Advanced Telescope for High-Energy Astrophysics (Athena) \citep{nandra2013hot}, and the enhanced X-ray Timing and Polarimetry (eXTP) \citep{zhang2019enhanced}, with sensitivity thresholds of \(10^{-17}, 3\times10^{-17},\) and \(3.3\times10^{-15} \rm \, erg\,s^{-1}\,cm^{-2}\) with integration time of  \(10^6\)s, respectively. By comparing the computed flux arriving at Earth with these thresholds, we can determine the number of sources detectable by each of these three observatories.

In addition to the requirement that the energy reaching Earth exceeds the detector threshold, the luminosity at the time of the outburst must also be greater than that of the host galaxy. Following the simulation by Dong-Páez et al. \cite{dong2023multi}, we adopt a probability of 30\% as the fraction of systems that satisfy these additional criteria. The final number of detectable sources for the three X-ray observatories is summarized in Table \ref{X-ray_number}.

\begin{table}
\setlength{\tabcolsep}{28pt}
\caption{\textbf{The table presents the number of sources from the three catalogs that can be detected by the three X-ray observatories, as well as the number of sources that can be jointly detected with LISA}: The number of sources exceeding the detector threshold at the time of the burst is multiplied by 30\% to simulate the number of sources that meet all detection criteria.}
\begin{tabular}{ccccc}
\hline
catalog & total number & eXTP & Athena & AXIS \\ \hline
pop\MakeUppercase{\romannumeral 3}  & 8735                                                   & 1.2                                                   & 8.7                                                    & 12.9                                                   \\
Q3-d    & 409                                                    & 7.20                                                   & 73.5                                                    & 92.1                                                  \\
Q3-nod  & 6122                                                   & 2.4                                                   & 21.9                                                     & 42.0                                                  \\ \hline
\end{tabular}
\label{X-ray_number}
\end{table}

After obtaining the catalog of sources that can be simultaneously detected in both electromagnetic and GWs, we can estimate their constraints on the speed of GWs. In Section \ref{introduction}, we provided the formula for constraining the GW speed as Eq.\ref{time_diff}. However, for the high redshift of SMBHBs, we have to take its impact into consideration \citep{nishizawa2014measuring}:
\begin{equation} 
\Delta t=(1+z)\Delta t_e-{\Delta c\over c}\int_0^z{\mathrm{d}z\over H(z)}
\end{equation} 
where $\Delta t_e$ is the difference in their emission times and $\Delta t$ is the time difference between the two signals. The total uncertainty in $\Delta t$ is composed of the timing uncertainties of both signals. Based on the simulation in the previous section, we can obtain the timing accuracy of the GW signal detected by LISA. For example, Figure \ref{cor1570} shows that it can be measured with an accuracy of around 0.1 seconds. Next, we estimate the timing accuracy for the electromagnetic signal.

Dong-Páez et al. \cite{dong2023multi} suggests that the brightening occurs within a timescale $\sim t_{\rm vis}$ before and after the merger \citep{armitage2002accretion,cerioli2016gas}, which corresponds to the time required for the cavity formed during the merger to be refilled \citep{milosavljevic2005afterglow}:
\begin{equation}
    t_{\rm vis} = 0.1(M_\bullet/10^6M_\odot)(\alpha_{\rm vis}/0.1)^{-8/5}(h_{\rm vis}/0.1)^{-16/5}\rm yr
\end{equation}
where $\alpha_{\rm vis}$ is the disc viscosity parameter and $h_{\rm vis}$ is the disc aspect ratio. This timescale can thus serve as an estimate of the timing precision associated with the electromagnetic counterpart. For the emission times difference $\Delta t_e$, we estimate its upper limit using the time it takes for light to travel across the outer radius of the black hole's accretion disk, which is approximately 75 times the Schwarzschild radius. 

It is clear that the uncertainty in the GW signal timing is significantly smaller than the uncertainties in the other two timing measurements. Therefore, in our calculation of constraints on the speed of GWs, we neglect this uncertainty.

\begin{figure}
    \centering
    \includegraphics[width=.8\linewidth]{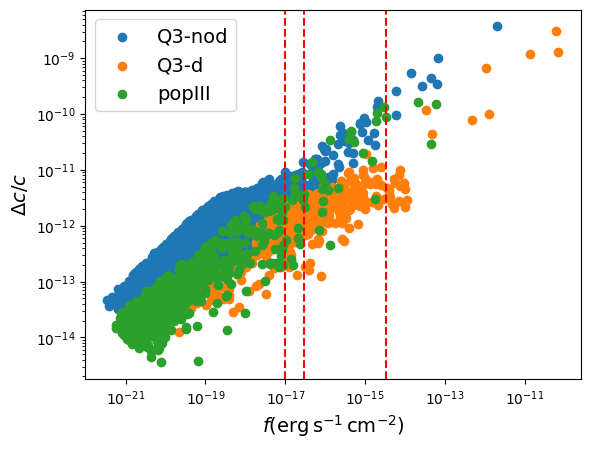}
    \includegraphics[width=.8\linewidth]{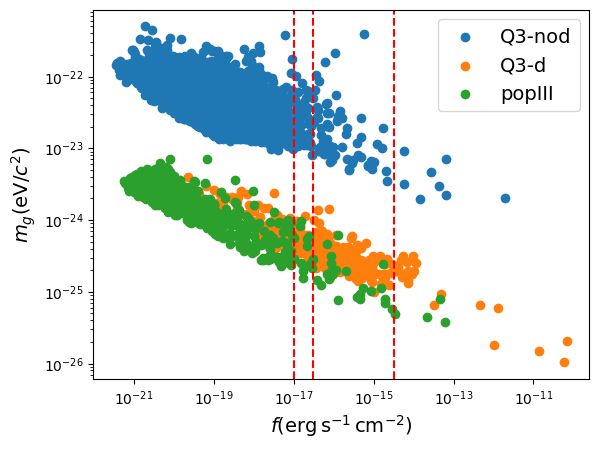}
    \caption{\textbf{The figure illustrates the relationship between the X-ray burst intensity at Earth from mergers in the three catalogs and the source's capability to constrain the speed of GWs and the graviton mass}: The three dashed lines represent the sensitivity thresholds of AXIS, Athena, and eXTP, respectively. Sources to the right of each dashed line are detectable by the corresponding X-ray telescope.}
    \label{L_c}
\end{figure}

Thus, using Equation \ref{time_diff}, we can determine the constraint on \(\Delta c/c\) for each source. We plot its relationship with the flux reaching Earth in the upper panel of Figure \ref{L_c}. We find that the two quantities exhibit a strong linear correlation in logarithmic space. The three dashed lines in the figure correspond to the sensitivity thresholds of AXIS, Athena, and eXTP. If the signals predicted by this model can be jointly detected by these instruments together with LISA, then the speed deviation \(\Delta c/c\) could be constrained to the level of \(10^{-13}-10^{-12}\), \(10^{-13}-2\times10^{-12}\), and \(2\times10^{-12}-10^{-10}\), respectively.

We can convert the constraint on \(\Delta c/c\) into a constraint on the graviton mass using the relationship:  
\begin{equation}
    \frac{\Delta c}{c}=\frac{1}{2}\frac{m_g^2c^4}{E^2}
\end{equation}  
where \(E=hf\) is the graviton's energy. By taking the frequency as the GW frequency at the merger peak, we can transform the velocity difference constraint into a graviton mass constraint, as shown in the lower panel of Figure \ref{L_c}. 

In the figure, we observe an inverse relationship between the energy reaching Earth and the graviton mass constraint—sources with higher received energy provide stronger constraints on the graviton mass. Different source catalogs can constrain the graviton mass to different extents. Among them, the Q3-d catalog is the best, reaching \(10^{-26} \mathrm{eV}/c^2\), while the Q3-nod catalog is the worst, with a limit of only \(10^{-24} \mathrm{eV}/c^2\).

\section{Summary}

In work, we applied full Bayesian analysis on simulated LISA to constrain the dispersive propagation GW effects.  Our conclusion is that LISA's observation on individual SMBHB mergers can be used to constrain the graviton mass \(m_g\) to the order of \(10^{-26} \, \mathrm{eV}/c^2\), which is three orders of magnitude better than the results from LIGO \citep{abbott2019tests}. Our estimation is in consistent with previous studies, which used simplified method of Fisher Information Matrix method \citep{gao2023testing}. We found that the upper limit of $m_g$ from individual source has a power-law dependence on both its SNR and chirp mass, with power indices -0.5 and -0.4 respectively. We also explore for the first time that, when there are two dispersive terms ($\mathbb{A}_{\alpha_1}p^{\alpha_1}+\mathbb{A}_{\alpha_2}p^{\alpha_2}$) in the dispersion relationship, how well the constraints on each coefficients (see results in Figure \ref{violin}). When we treat the dispersive index $\alpha$ as another unknown parameter to be inferred (among a series of discrete values), we found that the Beysian evidence is not the best indicator for the true $\alpha$. Instead, we propose another statistic $Q$ as a better indicator. 

Furthermore, by applying the population models of SMBHB mergers of K16, we simulate the samples of sources to be detected by LISA under different observation campaigns. We study the constraints on $m_g$, as long as other dispersive coefficients $\mathbb{A}_\alpha$ assuming a different dispersion index $\alpha$ from a sample of sources. The results of upper limits of  $\mathbb{A}_\alpha$ are listed in Table \ref{evidence}. We also compare our results with those using stellar-mass BBH merger GW events in GWTC-3. It is showed that, in cases where $\alpha<2$, the LISA observation on SMBHB merger population can results in much better constraints. 

By assuming a particular SMBHB EMW radiation model \citep{dong2023multi}, we calculate the X-ray flux from each sources in K16 population, and determine a sample of sources which can be jointly detected by LISA and X-ray telescopes. Using this multi-messenger catlaogue, the relative difference between the velocity of GW and that of EMW \(\Delta c/c\) can be constrained to \(10^{-13}\) (dispersion model-independent).  

The constraint on the GW speed is highly dependent on the sensitivity threshold of X-ray telescopes. As the threshold decreases and the detection capability of X-ray telescopes improves, the constraints derived from detected sources become more stringent. For the three telescopes, the expected constraints on \(\Delta c/c\) are in the ranges of \(10^{-13}-10^{-12}\), \(10^{-13}-2\times10^{-12}\), and \(2\times10^{-12}-10^{-10}\), respectively. Regarding the graviton mass, stronger signals provide better constraints. Therefore, lowering the telescope threshold does not significantly improve the final result. The graviton mass is expected to be constrained in the range of \(10^{-26}-10^{-24} \mathrm{eV}/c^2\).

\begin{acknowledgments}
This work is supported by the Chinese Academy of Sciences (Grant No. E329A3M1 and E3545KU2).
\end{acknowledgments}

\bibliography{apssamp}

@article{abbott2016observation,
  title={Observation of gravitational waves from a binary black hole merger},
  author={Abbott, Benjamin P and Abbott, Richard and Abbott, TDe and Abernathy, MR and Acernese, Fausto and Ackley, Kendall and Adams, Carl and Adams, Thomas and Addesso, Paolo and Adhikari, Rana X and others},
  journal={Physical review letters},
  volume={116},
  number={6},
  pages={061102},
  year={2016},
  publisher={APS}
}

@article{abbott2016tests,
  title={Tests of general relativity with GW150914},
  author={Abbott, Benjamin P and Abbott, R and Abbott, TD and Abernathy, MR and Acernese, Fausto and Ackley, K and Adams, C and Adams, T and Addesso, Paolo and Adhikari, RX and others},
  journal={Physical review letters},
  volume={116},
  number={22},
  pages={221101},
  year={2016},
  publisher={APS}
}

@article{abbott2017gw170817,
  title={GW170817: observation of gravitational waves from a binary neutron star inspiral},
  author={Abbott, Benjamin P and Abbott, Rich and Abbott, TDea and Acernese, Fausto and Ackley, Kendall and Adams, Carl and Adams, Thomas and Addesso, Paolo and Adhikari, Rana X and Adya, Vaishali B and others},
  journal={Physical review letters},
  volume={119},
  number={16},
  pages={161101},
  year={2017},
  publisher={APS}
}

@article{abbott2021tests,
  title={Tests of general relativity with GWTC-3},
  author={Abbott, R and Abe, H and Acernese, F and Ackley, K and Adhikari, N and Adhikari, RX and Adkins, VK and Adya, VB and Affeldt, C and Agarwal, D and others},
  journal={arXiv preprint arXiv:2112.06861},
  year={2021}
}

@article{abbott2019tests,
  title={Tests of general relativity with the binary black hole signals from the LIGO-Virgo catalog GWTC-1},
  author={Abbott, BP and Abbott, R and Abbott, TD and Abraham, S and Acernese, Fausto and Ackley, K and Adams, C and Adhikari, Rana X and Adya, VB and Affeldt, C and others},
  journal={Physical Review D},
  volume={100},
  number={10},
  pages={104036},
  year={2019},
  publisher={APS}
}

@article{amaro2017laser,
  title={Laser interferometer space antenna},
  author={Amaro-Seoane, Pau and Audley, Heather and Babak, Stanislav and Baker, John and Barausse, Enrico and Bender, Peter and Berti, Emanuele and Binetruy, Pierre and Born, Michael and Bortoluzzi, Daniele and others},
  journal={arXiv preprint arXiv:1702.00786},
  year={2017}
}

@article{ruan2020taiji,
  title={Taiji program: Gravitational-wave sources},
  author={Ruan, Wen-Hong and Guo, Zong-Kuan and Cai, Rong-Gen and Zhang, Yuan-Zhong},
  journal={International Journal of Modern Physics A},
  volume={35},
  number={17},
  pages={2050075},
  year={2020},
  publisher={World Scientific}
}

@article{mei2021tianqin,
  title={The TianQin project: Current progress on science and technology},
  author={Mei, Jianwei and Bai, Yan-Zheng and Bao, Jiahui and Barausse, Enrico and Cai, Lin and Canuto, Enrico and Cao, Bin and Chen, Wei-Ming and Chen, Yu and Ding, Yan-Wei and others},
  journal={Progress of Theoretical and Experimental Physics},
  volume={2021},
  number={5},
  pages={05A107},
  year={2021},
  publisher={Oxford University Press}
}

@article{agazie2023nanograv,
  title={The NANOGrav 15 yr data set: Evidence for a gravitational-wave background},
  author={Agazie, Gabriella and Anumarlapudi, Akash and Archibald, Anne M and Arzoumanian, Zaven and Baker, Paul T and B{\'e}csy, Bence and Blecha, Laura and Brazier, Adam and Brook, Paul R and Burke-Spolaor, Sarah and others},
  journal={The Astrophysical Journal Letters},
  volume={951},
  number={1},
  pages={L8},
  year={2023},
  publisher={IOP Publishing}
}

@article{antoniadis2023second,
  title={The second data release from the European Pulsar Timing Array-III. Search for gravitational wave signals},
  author={Antoniadis, J and Arumugam, P and Arumugam, S and Babak, S and Bagchi, M and Nielsen, A-S Bak and Bassa, CG and Bathula, A and Berthereau, A and Bonetti, M and others},
  journal={Astronomy \& Astrophysics},
  volume={678},
  pages={A50},
  year={2023},
  publisher={EDP Sciences}
}

@article{reardon2023search,
  title={Search for an isotropic gravitational-wave background with the Parkes Pulsar Timing Array},
  author={Reardon, Daniel J and Zic, Andrew and Shannon, Ryan M and Hobbs, George B and Bailes, Matthew and Di Marco, Valentina and Kapur, Agastya and Rogers, Axl F and Thrane, Eric and Askew, Jacob and others},
  journal={The Astrophysical Journal Letters},
  volume={951},
  number={1},
  pages={L6},
  year={2023},
  publisher={IOP Publishing}
}

@article{xu2023searching,
  title={Searching for the nano-Hertz stochastic gravitational wave background with the Chinese Pulsar Timing Array Data Release I},
  author={Xu, Heng and Chen, Siyuan and Guo, Yanjun and Jiang, Jinchen and Wang, Bojun and Xu, Jiangwei and Xue, Zihan and Caballero, R Nicolas and Yuan, Jianping and Xu, Yonghua and others},
  journal={Research in Astronomy and Astrophysics},
  volume={23},
  number={7},
  pages={075024},
  year={2023},
  publisher={IOP Publishing}
}

@article{gao2023testing,
  title={Testing alternative theories of gravity with space-based gravitational wave detectors},
  author={Gao, Qing and You, Yujie and Gong, Yungui and Zhang, Chao and Zhang, Chunyu},
  journal={Physical Review D},
  volume={108},
  number={2},
  pages={024027},
  year={2023},
  publisher={APS}
}

@article{klein2016science,
  title={Science with the space-based interferometer eLISA: Supermassive black hole binaries},
  author={Klein, Antoine and Barausse, Enrico and Sesana, Alberto and Petiteau, Antoine and Berti, Emanuele and Babak, Stanislav and Gair, Jonathan and Aoudia, Sofiane and Hinder, Ian and Ohme, Frank and others},
  journal={Physical Review D},
  volume={93},
  number={2},
  pages={024003},
  year={2016},
  publisher={APS}
}

@article{amelino2001testable,
  title={Testable scenario for relativity with minimum length},
  author={Amelino-Camelia, Giovanni},
  journal={Physics Letters B},
  volume={510},
  number={1-4},
  pages={255--263},
  year={2001},
  publisher={Elsevier}
}

@article{magueijo2002lorentz,
  title={Lorentz invariance with an invariant energy scale},
  author={Magueijo, Joao and Smolin, Lee},
  journal={Physical Review Letters},
  volume={88},
  number={19},
  pages={190403},
  year={2002},
  publisher={APS}
}

@article{amelino2002special,
  title={Special treatment},
  author={Amelino-Camelia, Giovanni},
  journal={Nature},
  volume={418},
  number={6893},
  pages={34--35},
  year={2002},
  publisher={Nature Publishing Group UK London}
}

@article{amelino2010doubly,
  title={Doubly-special relativity: Facts, myths and some key open issues},
  author={Amelino-Camelia, Giovanni},
  journal={Symmetry},
  volume={2},
  number={1},
  pages={230--271},
  year={2010},
  publisher={MDPI}
}

@article{sefiedgar2011modified,
  title={Modified dispersion relations in extra dimensions},
  author={Sefiedgar, AS and Nozari, K and Sepangi, HR},
  journal={Physics Letters B},
  volume={696},
  number={1-2},
  pages={119--123},
  year={2011},
  publisher={Elsevier}
}

@article{hovrava2009membranes,
  title={Membranes at quantum criticality},
  author={Ho{\v{r}}ava, Petr},
  journal={Journal of High Energy Physics},
  volume={2009},
  number={03},
  pages={020},
  year={2009},
  publisher={IOP Publishing}
}

@article{hovrava2009quantum,
  title={Quantum gravity at a Lifshitz point},
  author={Ho{\v{r}}ava, Petr},
  journal={Physical Review D—Particles, Fields, Gravitation, and Cosmology},
  volume={79},
  number={8},
  pages={084008},
  year={2009},
  publisher={APS}
}

@article{vacaru2012modified,
  title={Modified dispersion relations in Ho{\v{r}}ava--Lifshitz gravity and Finsler brane models},
  author={Vacaru, Sergiu I},
  journal={General Relativity and Gravitation},
  volume={44},
  pages={1015--1042},
  year={2012},
  publisher={Springer}
}

@inproceedings{garattini2012modified,
  title={Modified dispersion relations and noncommutative geometry lead to a finite Zero Point Energy},
  author={Garattini, Remo},
  booktitle={AIP Conference Proceedings 11},
  volume={1446},
  number={1},
  pages={298--310},
  year={2012},
  organization={American Institute of Physics}
}

@article{garattini2011modified,
  title={Modified dispersion relations lead to a finite zero point gravitational energy},
  author={Garattini, Remo and Mandanici, Gianluca},
  journal={Physical Review D—Particles, Fields, Gravitation, and Cosmology},
  volume={83},
  number={8},
  pages={084021},
  year={2011},
  publisher={APS}
}

@article{garattini2012particle,
  title={Particle propagation and effective space-time in Gravity’s Rainbow},
  author={Garattini, Remo and Mandanici, Gianluca},
  journal={Physical Review D—Particles, Fields, Gravitation, and Cosmology},
  volume={85},
  number={2},
  pages={023507},
  year={2012},
  publisher={APS}
}

@article{mirshekari2012constraining,
  title={Constraining Lorentz-violating, modified dispersion relations with gravitational waves},
  author={Mirshekari, Saeed and Yunes, Nicol{\'a}s and Will, Clifford M},
  journal={Physical Review D—Particles, Fields, Gravitation, and Cosmology},
  volume={85},
  number={2},
  pages={024041},
  year={2012},
  publisher={APS}
}

@software{michael_katz_2024_10930980,
  author       = {Michael Katz and
                  CChapmanbird and
                  Lorenzo Speri and
                  Nikolaos Karnesis and
                  Korsakova, Natalia},
  title        = {mikekatz04/LISAanalysistools: First main release.},
  month        = apr,
  year         = 2024,
  publisher    = {Zenodo},
  version      = {v1.0.3},
  doi          = {10.5281/zenodo.10930980},
  url          = {https://doi.org/10.5281/zenodo.10930980}
}

@software{michael_katz_2023_7705496,
  author       = {Michael Katz and
                  Nikolaos Karnesis and
                  Natalia Korsakova},
  title        = {mikekatz04/Eryn: first full release},
  month        = mar,
  year         = 2023,
  publisher    = {Zenodo},
  version      = {v1.0.0},
  doi          = {10.5281/zenodo.7705496},
  url          = {https://doi.org/10.5281/zenodo.7705496}
}

@article{yi2022gravitational,
  title={The Gravitational Wave Universe Toolbox-A software package to simulate observations of the gravitational wave universe with different detectors},
  author={Yi, Shu-Xu and Nelemans, Gijs and Brinkerink, Christiaan and Kostrzewa-Rutkowska, Zuzanna and Timmer, Sjoerd T and Stoppa, Fiorenzo and Rossi, Elena M and Zwart, Simon F Portegies},
  journal={Astronomy \& Astrophysics},
  volume={663},
  pages={A155},
  year={2022},
  publisher={EDP Sciences}
}

@article{abbott2017gravitational,
  title={Gravitational waves and gamma-rays from a binary neutron star merger: GW170817 and GRB 170817A},
  author={Abbott, Benjamin P and Abbott, Robert and Abbott, TD and Acernese, F and Ackley, K and Adams, C and Adams, T and Addesso, P and Adhikari, RX and Adya, VB and others},
  journal={The Astrophysical Journal Letters},
  volume={848},
  number={2},
  pages={L13},
  year={2017},
  publisher={IOP Publishing}
}

@article{cao2024constraining,
  title={Constraining gravitational wave velocities using gravitational and electromagnetic wave observations of white dwarf binaries},
  author={Cao, Tian-Yong and Kumar, Ankit and Yi, Shu-Xu},
  journal={Monthly Notices of the Royal Astronomical Society},
  volume={533},
  number={1},
  pages={551--560},
  year={2024},
  publisher={Oxford University Press}
}

@article{Kostelecky2016Testing,
  title={Testing local Lorentz invariance with gravitational waves},
  author={Kosteleck{\`y}, V Alan and Mewes, Matthew},
  journal={Physics Letters B},
  volume={757},
  pages={510--514},
  year={2016},
  publisher={Elsevier}
}

@article{Mewes2019Signals,
  title={Signals for Lorentz violation in gravitational waves},
  author={Mewes, Matthew},
  journal={Physical Review D},
  volume={99},
  number={10},
  pages={104062},
  year={2019},
  publisher={APS}
}

@article{rao2024simulation,
  title={Simulation Study on Constraining Gravitational Wave Propagation Speed by Gravitational Wave and Gamma-ray Burst Joint Observation on Binary Neutron Star Mergers},
  author={Rao, Jin-Hui and Yi, Shu-Xu and Tao, Lian and Tang, Qing-Wen},
  journal={Research in Astronomy and Astrophysics},
  volume={24},
  number={8},
  pages={085004},
  year={2024},
  publisher={IOP Publishing}
}

@article{will2014confrontation,
  title={The confrontation between general relativity and experiment},
  author={Will, Clifford M},
  journal={Living reviews in relativity},
  volume={17},
  pages={1--117},
  year={2014},
  publisher={Springer}
}

@article{kostelecky2002signals,
  title={Signals for Lorentz violation in electrodynamics},
  author={Kosteleck{\`y}, V Alan and Mewes, Matthew},
  journal={Physical Review D},
  volume={66},
  number={5},
  pages={056005},
  year={2002},
  publisher={APS}
}

@article{colladay1997cpt,
  title={CPT violation and the standard model},
  author={Colladay, Don and Kosteleck{\`y}, V Alan},
  journal={Physical Review D},
  volume={55},
  number={11},
  pages={6760},
  year={1997},
  publisher={APS}
}

@article{colladay1998lorentz,
  title={Lorentz-violating extension of the standard model},
  author={Colladay, Don and Kosteleck{\`y}, V Alan},
  journal={Physical Review D},
  volume={58},
  number={11},
  pages={116002},
  year={1998},
  publisher={APS}
}

@article{volonteri2003assembly,
  title={The assembly and merging history of supermassive black holes in hierarchical models of galaxy formation},
  author={Volonteri, Marta and Haardt, Francesco and Madau, Piero},
  journal={The Astrophysical Journal},
  volume={582},
  number={2},
  pages={559},
  year={2003},
  publisher={IOP Publishing}
}

@article{shen2009supermassive,
  title={Supermassive black holes in the hierarchical universe: a general framework and observational tests},
  author={Shen, Yue},
  journal={The Astrophysical Journal},
  volume={704},
  number={1},
  pages={89},
  year={2009},
  publisher={IOP Publishing}
}

@article{dong2023multi,
  title={Multi-messenger study of merging massive black holes in the OBELISK simulation: Gravitational waves, electromagnetic counterparts, and their link to galaxy and black-hole populations},
  author={Dong-P{\'a}ez, Chi An and Volonteri, Marta and Beckmann, Ricarda S and Dubois, Yohan and Mangiagli, Alberto and Trebitsch, Maxime and Vergani, Susanna D and Webb, Natalie A},
  journal={Astronomy \& Astrophysics},
  volume={676},
  pages={A2},
  year={2023},
  publisher={EDP Sciences}
}

@article{gutierrez2022electromagnetic,
  title={Electromagnetic signatures from supermassive binary black holes approaching merger},
  author={Guti{\'e}rrez, Eduardo M and Combi, Luciano and Noble, Scott C and Campanelli, Manuela and Krolik, Julian H and Armengol, Federico L{\'o}pez and Garc{\'\i}a, Federico},
  journal={The Astrophysical Journal},
  volume={928},
  number={2},
  pages={137},
  year={2022},
  publisher={IOP Publishing}
}

@article{morrison1983interstellar,
  title={Interstellar photoelectric absorption cross sections, 0.03-10 keV},
  author={Morrison, Robert and McCammon, Dan},
  journal={Astrophysical Journal, Part 1 (ISSN 0004-637X), vol. 270, July 1, 1983, p. 119-122.},
  volume={270},
  pages={119--122},
  year={1983}
}

@inproceedings{mushotzky2018axis,
  title={AXIS: a probe class next generation high angular resolution x-ray imaging satellite},
  author={Mushotzky, R},
  booktitle={Space Telescopes and Instrumentation 2018: Ultraviolet to Gamma Ray},
  volume={10699},
  pages={570--591},
  year={2018},
  organization={SPIE}
}

@article{nandra2013hot,
  title={The Hot and Energetic Universe: A White Paper presenting the science theme motivating the Athena+ mission},
  author={Nandra, Kirpal and Barret, Didier and Barcons, Xavier and Fabian, Andy and Herder, Jan-Willem den and Piro, Luigi and Watson, Mike and Adami, Christophe and Aird, James and Afonso, Jose Manuel and others},
  journal={arXiv preprint arXiv:1306.2307},
  year={2013}
}

@article{zhang2019enhanced,
  title={The enhanced X-ray Timing and Polarimetry mission—eXTP},
  author={Zhang, ShuangNan and Santangelo, Andrea and Feroci, Marco and Xu, YuPeng and Lu, FangJun and Chen, Yong and Feng, Hua and Zhang, Shu and Brandt, S{\o}ren and Hernanz, Margarita and others},
  journal={Science China Physics, Mechanics \& Astronomy},
  volume={62},
  pages={1--25},
  year={2019},
  publisher={Springer}
}

@article{arun2022new,
  title={New horizons for fundamental physics with LISA},
  author={Arun, KG and Belgacem, Enis and Benkel, Robert and Bernard, Laura and Berti, Emanuele and Bertone, Gianfranco and Besancon, Marc and Blas, Diego and B{\"o}hmer, Christian G and Brito, Richard and others},
  journal={Living Reviews in Relativity},
  volume={25},
  number={1},
  pages={4},
  year={2022},
  publisher={Springer}
}

@article{cornish2021heterodyned,
  title={Heterodyned likelihood for rapid gravitational wave parameter inference},
  author={Cornish, Neil J},
  journal={Physical Review D},
  volume={104},
  number={10},
  pages={104054},
  year={2021},
  publisher={APS}
}

@article{zackay2018relative,
  title={Relative binning and fast likelihood evaluation for gravitational wave parameter estimation},
  author={Zackay, Barak and Dai, Liang and Venumadhav, Tejaswi},
  journal={arXiv preprint arXiv:1806.08792},
  year={2018}
}

@article{novikov1973astrophysics,
  title={Astrophysics of black holes},
  author={Novikov, Igor D and Thorne, Kip S},
  journal={Black holes (Les astres occlus)},
  volume={1},
  pages={343--450},
  year={1973}
}

@article{noble2012circumbinary,
  title={Circumbinary magnetohydrodynamic accretion into inspiraling binary black holes},
  author={Noble, Scott C and Mundim, Bruno C and Nakano, Hiroyuki and Krolik, Julian H and Campanelli, Manuela and Zlochower, Yosef and Yunes, Nicol{\'a}s},
  journal={The Astrophysical Journal},
  volume={755},
  number={1},
  pages={51},
  year={2012},
  publisher={IOP Publishing}
}

@article{shi2012three,
  title={Three-dimensional magnetohydrodynamic simulations of circumbinary accretion disks: disk structures and angular momentum transport},
  author={Shi, Ji-Ming and Krolik, Julian H and Lubow, Stephen H and Hawley, John F},
  journal={The Astrophysical Journal},
  volume={749},
  number={2},
  pages={118},
  year={2012},
  publisher={IOP Publishing}
}

@article{shi2015three,
  title={Three-dimensional MHD simulation of circumbinary accretion disks. II. Net accretion rate},
  author={Shi, Ji-Ming and Krolik, Julian H},
  journal={The Astrophysical Journal},
  volume={807},
  number={2},
  pages={131},
  year={2015},
  publisher={IOP Publishing}
}

@article{cerioli2016gas,
  title={Gas squeezing during the merger of a supermassive black hole binary},
  author={Cerioli, Alice and Lodato, Giuseppe and Price, Daniel J},
  journal={Monthly Notices of the Royal Astronomical Society},
  volume={457},
  number={1},
  pages={939--948},
  year={2016},
  publisher={Oxford University Press}
}

@article{armitage2002accretion,
  title={Accretion during the merger of supermassive black holes},
  author={Armitage, Philip J and Natarajan, Priyamvada},
  journal={The Astrophysical Journal},
  volume={567},
  number={1},
  pages={L9},
  year={2002},
  publisher={IOP Publishing}
}

@article{milosavljevic2005afterglow,
  title={The afterglow of massive black hole coalescence},
  author={Milosavljevi{\'c}, Milo{\v{s}} and Phinney, E Sterl},
  journal={The Astrophysical Journal},
  volume={622},
  number={2},
  pages={L93},
  year={2005},
  publisher={IOP Publishing}
}

@article{nishizawa2014measuring,
  title={Measuring speed of gravitational waves by observations of photons and neutrinos from compact binary mergers and supernovae},
  author={Nishizawa, Atsushi and Nakamura, Takashi},
  journal={Physical Review D},
  volume={90},
  number={4},
  pages={044048},
  year={2014},
  publisher={APS}
}

@article{cutler1994gravitational,
  title={Gravitational waves from merging compact binaries: How accurately can one extract the binary’s parameters from the inspiral waveform?},
  author={Cutler, Curt and Flanagan, Eanna E},
  journal={Physical Review D},
  volume={49},
  number={6},
  pages={2658},
  year={1994},
  publisher={APS}
}

@article{london2018first,
  title={First higher-multipole model of gravitational waves from spinning and coalescing black-hole binaries},
  author={London, Lionel and Khan, Sebastian and Fauchon-Jones, Edward and Garc{\'\i}a, Cecilio and Hannam, Mark and Husa, Sascha and Jim{\'e}nez-Forteza, Xisco and Kalaghatgi, Chinmay and Ohme, Frank and Pannarale, Francesco},
  journal={Physical review letters},
  volume={120},
  number={16},
  pages={161102},
  year={2018},
  publisher={APS}
}

@article{husa2016frequency,
  title={Frequency-domain gravitational waves from nonprecessing black-hole binaries. I. New numerical waveforms and anatomy of the signal},
  author={Husa, Sascha and Khan, Sebastian and Hannam, Mark and P{\"u}rrer, Michael and Ohme, Frank and Forteza, Xisco Jim{\'e}nez and Boh{\'e}, Alejandro},
  journal={Physical Review D},
  volume={93},
  number={4},
  pages={044006},
  year={2016},
  publisher={APS}
}

@article{khan2016frequency,
  title={Frequency-domain gravitational waves from nonprecessing black-hole binaries. II. A phenomenological model for the advanced detector era},
  author={Khan, Sebastian and Husa, Sascha and Hannam, Mark and Ohme, Frank and P{\"u}rrer, Michael and Forteza, Xisco Jim{\'e}nez and Boh{\'e}, Alejandro},
  journal={Physical Review D},
  volume={93},
  number={4},
  pages={044007},
  year={2016},
  publisher={APS}
}

@article{marsat2018fourier,
  title={Fourier-domain modulations and delays of gravitational-wave signals},
  author={Marsat, Sylvain and Baker, John G},
  journal={arXiv preprint arXiv:1806.10734},
  year={2018}
}

@article{marsat2021exploring,
  title={Exploring the Bayesian parameter estimation of binary black holes with LISA},
  author={Marsat, Sylvain and Baker, John G and Canton, Tito Dal},
  journal={Physical Review D},
  volume={103},
  number={8},
  pages={083011},
  year={2021},
  publisher={APS}
}

@article{babak2021lisa,
  title={LISA sensitivity and SNR calculations},
  author={Babak, Stanislav and Hewitson, Martin and Petiteau, Antoine},
  journal={arXiv preprint arXiv:2108.01167},
  year={2021}
}

@article{lartillot2006computing,
  title={Computing Bayes factors using thermodynamic integration},
  author={Lartillot, Nicolas and Philippe, Herv{\'e}},
  journal={Systematic biology},
  volume={55},
  number={2},
  pages={195--207},
  year={2006},
  publisher={Oxford University Press}
}

\end{document}